\documentclass[aps,twocolumn,amsmath,amssymb,showpacs]{revtex4-1}
\usepackage{graphicx}
\usepackage{dcolumn}
\usepackage{bm}

\usepackage[caption=false]{subfig}
\usepackage{latexsym}
\usepackage{amsthm}
\usepackage{amsmath}
\usepackage{braket}
\usepackage{mathtools}
\usepackage[T1]{fontenc}
\allowdisplaybreaks

\usepackage[top = 1. in,bottom = 0.9 in, left = 1 in, right=1 in]{geometry}

 \newcommand{\arXiv}[1]{\href{http://www.arXiv.org/abs/#1}{arXiv:#1}}
\usepackage[colorlinks=true, linkcolor=blue, bookmarks=true]{hyperref}

\makeatletter
\renewcommand\section{\@startsection {section}{1}{\z@}%
                  {-3.5ex \@plus -1ex \@minus -.2ex}
                  {2.3ex \@plus.2ex}%
                  {\normalfont\bfseries}}
\renewcommand\subsection{\@startsection{subsection}{2}{\z@}%
                   {-3.25ex\@plus -1ex \@minus -.2ex}%
                   {1.5ex \@plus .2ex}%
                   {\normalfont\bfseries}}
\makeatother

\newcommand{\al}{\alpha}
\newcommand{\ad}{\alpha^\dagger}

\newcommand{\beq}{\begin{equation}}
\newcommand{\eeq}{\end{equation}}
\newcommand{\ber}{\begin{array}}
\newcommand{\eer}{\end{array}}
\newcommand{\del}{\partial}
\newcommand{\sssty}{\scriptscriptstyle}

\newcommand{\s}{\sigma}
\newcommand{\te}{\theta}

\newcommand{\de}{\delta}

\newcommand{\eps}{\varepsilon}

\newcommand{\ena}{\end{eqnarray}}
\newcommand{\beqa}{\begin{eqnarray}}
\newcommand{\eeqa}{\end{eqnarray}}
\newcommand{\bea}{\begin{eqnarray}}
\newcommand{\eea}{\end{eqnarray}}

\newcommand{\bfr}{{\bf r}}
\newcommand{\bfR}{{\bf R}}
\newcommand{\bfs}{{\bf s}}
\newcommand{\dr}{d\bfr}
\newcommand{\dR}{d\bfR}
\newcommand{\ds}{d\bfs}

\theoremstyle{remark}

\begin{document}

\title{Energy level splitting for weakly interacting bosons in a harmonic trap}
\author {Ben Craps,$^{1}$ Marine De Clerck,$^{1}$ Oleg Evnin$^{2,1}$ and Surbhi Khetrapal$^{1}$\vspace{2mm}}

\affiliation{ $^{1}$Theoretische Natuurkunde, Vrije Universiteit Brussel (VUB) and\\
	The International Solvay Institutes, Pleinlaan 2, B-1050 Brussels, Belgium\vspace{1mm}\\
$^{2}$Department of Physics, Faculty of Science, Chulalongkorn University, Phayathai Rd., Bangkok 10330, Thailand}

\begin{abstract}
We consider identical quantum bosons with weak contact interactions in a two-dimensional isotropic harmonic trap. When the interactions are turned off, the energy levels are equidistant and highly degenerate. At linear order in the coupling parameter, these degenerate levels split, and we study the patterns of this splitting. It turns out that the problem is mathematically identical to diagonalizing the quantum resonant system of the two-dimensional Gross-Pitaevskii equation, whose classical counterpart has been previously studied in the mathematical literature on turbulence. Our purpose is to explore the implications of the symmetries and energy bounds of this resonant system, previously studied for the classical case, for the quantum level splitting. Simplifications in computing the splitting spectrum numerically result from exploiting the symmetries. The highest energy state emanating from each unperturbed level is explicitly described by our analytics. We furthermore discuss the energy level spacing distributions in the spirit of quantum chaos theory. After separating the eigenvalues into blocks with respect to the known conservation laws, we observe the Wigner-Dyson statistics within specific large blocks, which leaves little room for further integrable structures in the problem beyond the symmetries that are already explicitly known.

\end{abstract}

\maketitle

\section{Introduction}

Energy levels of quantum identical interacting bosons in harmonic traps have often been studied in the literature \cite{split1,split2,split3,split4,split5,split6,split7,split8,split9,breathing}, motivated in particular by the physics of cold atomic gases.
Contact interactions between the bosons commonly appear in such studies as the simplest possible choice for 
the two-particle potential that is expected to retain realistic features. Our purpose in this article is to report on rich mathematical
structures emerging from studies of bosons with weak contact interactions in a two-dimensional isotropic harmonic trap, and the consequences
these structures have for the makeup of the energy spectrum.

Most of the past publications dealing with related systems have focused on particular small numbers of bosons. For instance, 
\cite{split1} reports an exact solution of the problem involving two bosons in a harmonic trap with contact interactions of arbitrary strength. A notable exception is \cite{breathing}, which studies
effects of symmetries on the spectrum of an arbitrary number of atoms with finite strength contact interactions in a harmonic trap. Our perspective will be quite similar, but the strength of the interactions will be assumed small (more precisely, our results are valid at linear order in the coupling parameter). This allows for more extensive analytic exploration of the structure of the spectrum.

If the interactions are turned off altogether, the eigenstates of the multi-boson system are simply constructed by having the bosons independently occupy the energy levels of the harmonic potential, and the total energy is a sum of contributions of the individual noninteracting bosons. Since the energies of harmonic potential eigenstates are integer in appropriate units, one ends up with integer energy eigenvalues for the multi-boson system, and the degeneracies of the energy levels are very high, since there are many ways to partition a given integer amount of energy into the integer energies of individual bosons. (These level degeneracies furthermore grow without bound as the energy is increased.) When weak contact interactions between the bosons are turned on, the highly degenerate free-boson energy levels split, and the pattern of the splitting at leading order in the interaction strength is computed by diagonalizing the interaction Hamiltonian within each degenerate unperturbed energy level, which is a standard quantum-mechanical procedure. These level splitting patterns will be the main subject of our analysis.

If one is interested in the splitting pattern of a given unperturbed level, diagonalizing a finite-sized matrix of the interaction Hamiltonian matrix elements is all one has to do. This is straightforward for numerical evaluation, yet little can be said about the general structure of the spectrum, since one matrix has to be diagonalized for each unperturbed level, and the sizes of the matrices (determined by the unperturbed level degeneracies) grow without bound as one moves to higher energies. However, specifically for bosons with contact interactions in an isotropic two-dimensional harmonic trap, the diagonalization problems involved display rich symmetry structures that relate the energy shifts of different unperturbed levels, and impose significant constraints on the patterns in the spectrum.

A key observation that makes our analysis possible is that the diagonalization problem involved in finding the energy shifts is identical to diagonalizing the quantized version of the resonant system of the two-dimensional Gross-Pitaevskii equation in an isotropic harmonic trap. The Gross-Pitaevskii equation (which physically describes the condensed regime of our interacting bosons) is a partial differential equation (PDE) that can be approximated in the weakly nonlinear regime by the corresponding resonant Hamiltonian system, an approach commonly taken in the mathematical literature on turbulence \cite{FGH,continuous}. The enhanced symmetries of this classical resonant system have been presented explicitly in the past, see \cite{continuous}. Quantum resonant systems, on the other hand, have been considered in \cite{quantres} from a perspective geared toward quantum chaos studies. Our strategy here is to translate the classical symmetries and energy bounds of \cite{continuous} to the corresponding quantum resonant system, and to relate the results to the energy level splitting of bosons with weak contact interactions.

Another issue that naturally comes to mind is whether there are symmetries in the problem beyond the ones we use. To shed light on this question, we turn to quantum chaos theory \cite{GMW,haake,DKPR}, which purports that integrable and chaotic systems are characterized by qualitatively different distributions of distances between neighboring energy levels. For integrable systems, the distribution is Poissonian \cite{btint}, so that, as far as the level spacing statistics is concerned, the energy levels appear random and completely uncorrelated. Chaotic systems, on the other hand, are associated with a phenomenon known as level repulsion, captured by the Wigner-Dyson distribution of the level spacings \cite{BGS}. In our case, we already know a few explicit operators that commute with the Hamiltonian, and one must factor this knowledge into the study of the level spacing statistics. If the statistics is built indiscriminately for all energy levels, one observes the Poisson distribution typical of integrable systems. This is merely a reflection of the fact that eigenvalues split into blocks with respect to the conserved quantities, and without correlations between eigenvalues in the different blocks, the joint level spacing statistics mimics that of random energy levels \cite{DKPR}. What is more relevant is to plot the level spacing distributions separately within such blocks. Once we do that taking into account all the conserved quantitites of \cite{continuous}, most of the large blocks display Wigner-Dyson-like, rather than Poissonian, distributions. This suggests that no integrability is to be expected, and further analytic structures, if any, must be very subtle. (We do, however, find Poisson-like distributions in a very restricted part of the spectrum characterized by rapid rotation.)

The paper is organized as follows: In section II, we review bosons with contact interactions in harmonic traps, and their energy level splitting at linear order in the coupling parameter, and then explain the connection between these considerations and quantum resonant systems. In section III, we demonstrate the use of our techniques by focusing on specific multiplets in the so-called Lowest Landau Level (LLL) sector, for which the analysis simplifies. In section IV, we generalize the analysis of section III to arbitrary energy levels. We conclude with a summary and discussion.


\section{Trapped bosons, energy\\ shifts and resonant systems}

We consider the second-quantized representation of identical bosons with contact interactions of strength $g$ in an isotropic harmonic potential:
\begin{align}
 &\mathcal{H} =\mathcal{H}_0 + g \,\mathcal{H}_{int},  \label{Horg}\\
&\mathcal{H}_0= \frac{1}{2}\int (  \nabla \Psi^\dagger \cdot \nabla \Psi + (x^2 + y^2)\Psi^\dagger \Psi)\,dx\,dy,\\ 
&\mathcal{H}_{int}=\pi \int \,  \Psi^{\dagger2} \Psi^{2} \, dx\,dy.\label{Hint}
\end{align}
Here, $\Psi(x,y)$ is a quantum nonrelativistic bosonic field satisfying the commutation relations
\beq
[\Psi^\dagger(x,y),\Psi(x',y')]=-\de(x-x')\,\de(y-y').
\eeq
We shall be focusing on small values of the coupling parameter, $g\ll 1$. It will be convenient to further assume $g\ge 0$ to simplify the wording (for example, to refer to the highest rather than lowest energy state within each fine structure multiplet), though our derivations are equally valid for small negative couplings. The factor of $\pi$ is inserted on the right-hand side of (\ref{Hint}) for future convenience and may be absorbed, if desired, into a redefinition of $g$.

We note that known subtleties exist with defining the operator product at the same point in (\ref{Hint}), but these subtleties will not be relevant for our treatment. The issue is commonly stated in the language of first quantization, where the naive product given in (\ref{Hint}) corresponds to the interparticle potential $V(\bfr_i-\bfr_j)\sim\de(\bfr_i-\bfr_j)$ between particles number $i$ and $j$ of an $N$-body system. It is known that wavefunctions of multiparticle systems with contact interactions possess singularities at coincident particle positions $\bfr_i=\bfr_j$, originating from the singularities of the Laplacian Green's function in more than one spacial dimension. Since one has to define products of the wave functions and the potential to formulate the Schr\"odinger equation, naive $\de$-functions in the potential are not acceptable and need to be replaced by a modification that can be multiplied by wavefunctions with specific singularities at coincident particle positions. This is clearly explained in \cite{split1} for two particles in three spatial dimensions, where the wavefunctions are allowed to have $1/r$ singularities and the $\de(\bfr)$ has to be replaced by the operator $\de(\bfr)\del_r r$. Note that, if applied to any nonsingular function, this operator acts exactly like a naive $\de$-function, but its action is also defined for functions with $1/r$ singularities. This also explains why this subtlety is irrelevant for our considerations: as we shall deal with the first order of perturbation theory in $g$, we only need to compute the matrix elements of the interaction Hamiltonian between noninteracting eigenstates. But the noninteracting eigenstates are smooth functions, without any singularities at coincident particle positions, and hence using naive $\de$-functions, and naive products of second-quantized fields in (\ref{Hint}), is just as good as using the correct regulated expressions, and evidently we shall not encounter any divergences in evaluating matrix elements of the naive expression (\ref{Hint}).

Setting for a moment $g=0$ in (\ref{Horg}), one obtains noninteracting bosons in a harmonic trap, and the energy spectrum descends directly from the one-particle spectrum by a simple addition of energies. One can decompose $\Psi$ in terms of the harmonic oscillator eigenfunctions $\psi_{nm}$ as
\beq
\Psi(x,y)=\sum_{n,m} \alpha_{nm} \psi_{nm}(x,y),
\label{psidecomp}
\eeq
with $\psi_{nm}$ satisfying
\begin{align}
\frac{1}{2}(-&\del_x^2- \del_y^2 + x^2 + y^2)\psi_{nm}=(n+1)\psi_{nm},\nonumber\\
&-i\del_\varphi \psi_{nm}=m\psi_{nm},
\end{align}
where $\varphi$ is the polar angle in the $(x,y)$-plane. These states thus carry $n+1$ units of energy and $m$ units of angular momentum, with $n$ being a nonnegative integer and $m \in \{-n, -n+2, \dots, n-2,n\}$. The creation-annihilation operators $\ad_{nm}$ and $\al_{nm}$ satisfy the standard commutation relations
\beq
[\ad_{nm},\al_{n'm'}]=-\de_{nn'}\de_{mm'}.
\eeq
The noninteracting Hamiltonian $\mathcal{H}_0$ is then expressed as
\beq
\mathcal{H}_0=\sum_{nm} n\, \ad_{nm} \al_{nm},
\label{H0al}
\eeq
where we have subtracted the irrelevant vacuum energy contribution by replacing $n+1$ with $n$. The eigenstates of $\mathcal{H}_0$ are simply given by the Fock basis states, generated by acting on the vacuum state $|0\rangle$ (one has $\al_{nm}|0\rangle=0$) with the creation operators $\ad_{nm}$ to produce a state with the set of occupation numbers $\{\eta_{nm}\}$, one for each one-particle mode labelled by $n$ and $m$:
\beq
|\{\eta_{nm}\}\rangle=\prod_{nm}\frac{( \ad_{nm} )^{\eta_{nm}}}{\sqrt{\eta_{nm}!}}|0\rangle.
\label{Fockdef}
\eeq
These satisfy, for any $n$ and $m$,
\beq
\ad_{nm} \al_{nm}|\{\eta\}\rangle=\eta_{nm} |\{\eta\}\rangle,
\eeq
and hence they are eigenstates of (\ref{H0al}) with eigenvalues
\beq
E_{\{\eta\}}=\sum_{nm} n \,\eta_{nm}.
\label{E0def}
\eeq
There are of course many ways to generate the same value of $E$ by combining the integer numbers $n$ and $\eta_{nm}$. Correspondingly, the energy levels of  (\ref{H0al}) are highly degenerate (and the degeneracies grow without bound as one moves to higher energies). 

A note is in order on the set of eigenfunctions to be used in the decomposition (\ref{psidecomp}). A common choice for the orthonormal set of two-dimensional harmonic oscillator energy eigenfunctions is given (in polar coordinates) by
\begin{equation}
\psi^{\mbox{\tiny norm.}}_{nm}= \sqrt{\frac{(\frac12(n-|m| ))!}{(\frac12(n+|m| ))!}} \frac{r^{|m|}}{\sqrt{\pi}} L^{|m|}_{\frac{n-|m|}{2}}(r^2)e^{-r^2/2}e^{im\phi},
\label{psipos}
\end{equation}
see, e.g., \cite{dahl}. 
Here, $L_n^{\alpha}$ are the generalized Laguerre polynomials. These functions are, of course, defined up to phase factors, which are completely irrelevant in the noninteracting theory. When considering interactions, however, different choices of the phase factors in (\ref{psipos}) will lead to equivalent theories, but may be more or less efficient from the standpoint of simplifying the algebraic expressions and making their structure more apparent. For the purposes of our algebra
it turns out beneficial to introduce an extra sign factor in the definition of the eigenfunctions as follows:
\beq
\psi_{nm}= (-1)^{\frac{1}{2}(m-|m|)}\psi^{\mbox{\tiny norm.}}_{nm}.
\eeq
With this sign factor, the wavefunctions can be conveniently transformed in a manner identical to the considerations of \cite{BBCE2} using identities for Laguerre polynomials. Using the explicit expressions for the Laguerre
polynomials,
\begin{equation}
L^{\mu}_n(\rho) = \sum^{n}_{k=0} (-1)^k \frac{(n+\mu)!}{k!(n-k)!(k+\mu)!}\rho^k,
\end{equation}
and remembering that factorials of negative numbers are infinite, we obtain for every integer $\mu$
\beq
L^{\mu}_n(\rho) = (-1)^{\mu} \frac{(n+\mu)!}{n!} \rho^{-\mu} L^{-\mu}_{\mu+n}(\rho). 
\eeq
Therefore, one can see that introducing the extra sign factors amounts to using eigenfunctions that can be written as the original $\psi^{\mbox{\tiny norm.}}_{nm}$ but without absolute values,
\begin{align}
&\sqrt{\frac{((n-|m| )/2)!}{((n+|m| )/2)!}} r^{|m|} L^{|m|}_{\frac{n-|m|}{2}}(r^2)e^{-r^2/2} \\
&= (-1)^{\frac{1}{2}(m-|m|)} \sqrt{\frac{((n-m )/2)!}{((n+m )/2)!}} r^{m} L^{m}_{\frac{n-m}{2}}(r^2)e^{-r^2/2},\nonumber
\end{align}
or
\begin{equation}
\psi_{nm}= \sqrt{\frac{(\frac12(n-m ))!}{(\frac12(n+m ))!}}\frac{r^{m}}{\sqrt{\pi}} L^{m}_{\frac{n-m}{2}}(r^2)e^{-r^2/2}e^{im\phi}.
\label{psipossgn}
\end{equation}
In the following, we shall use these expressions in the decompositions (\ref{psidecomp}), which in fact brings us in accord with the conventions of \cite{continuous} and lets us conveniently reuse the mathematical structures developed there.

We are now in a position to study weak contact interactions. To this end, we substitute the decomposition (\ref{psidecomp}) in the interaction Hamiltonian (\ref{Hint}) to obtain
\begin{align}
\mathcal{H}_{int}=&\hspace{2mm}{\textstyle\frac12} \,\sum_{\mathclap{\substack{n_1,n_2,n_3,n_4 \geq 0\\m_1+m_2 = m_3+ m_4}}} \hspace{2mm}C_{n_1n_2n_3n_4}^{m_1m_2m_3m_4}  {\alpha}^{\dagger}_{n_1m_1}{\alpha}^{\dagger}_{n_2m_2} {\alpha}_{n_3m_3}{\alpha}_{n_4m_4}.
\label{eq: interaction Hamiltonian}
\end{align}
The sum only contains terms satisfying $m_1+m_2 = m_3+ m_4$ as an immediate consequence of the angular momentum conservation by contact interactions. We have furthermore introduced the \emph{interaction coefficients} defined by
\begin{equation}
C_{n_1n_2n_3n_4}^{m_1m_2m_3m_4} =2\pi \int \psi_{n_1m_1}^* \psi_{n_2m_2}^*
\psi_{n_3m_3}
\psi_{n_4m_4} r \, dr \, d\phi
\label{intcoeff}
\end{equation}
whose properties will play a key role in our analysis. Note that, with the conventions we have adopted, 
\beq
C^{0000}_{0000}=1.
\label{C0000}
\eeq

A standard approach to small perturbations of quantum dynamics in confining potentials, known as the Rayleigh-Schr\"odinger perturbation theory, is to analyze the corrections to energy levels and eigenstates as power series in the interaction strength $g$. If the unperturbed levels are degenerate, as they are in the case at hand, level splitting will be in general induced by perturbations. At linear order in $g$, this level splitting is analyzed by computing the eigenvalues $\eps_I$ of the matrix
\beq
\langle \{\eta\}|\mathcal{H}_{int}| \{\eta'\}\rangle,
\label{blockdef}
\eeq 
where $| \{\eta\}\rangle$ and $| \{\eta'\}\rangle$ are two unperturbed Fock states of the form (\ref{Fockdef}) with the same value of the unperturbed energy given by (\ref{E0def}). Since $\mathcal{H}_{int}$ conserves the number of particles, $| \{\eta\}\rangle$ and $| \{\eta'\}\rangle$ must also contain the same total number of particles
\beq
N_{\{\eta\}}=\sum_{nm} \eta_{nm}.
\eeq
For any given $N$ and $E$, (\ref{blockdef}) is a finite-sized numerical matrix with the entries expressed through the interaction coefficients (\ref{intcoeff}). Once its eigenvalues $\eps_I$ have been found, the corresponding perturbed energy levels are given at order $g$ by
\beq
\tilde E_I=E+g\,\eps_I.
\label{enshifts}
\eeq

We note in passing that the structure of the unperturbed levels, and the diagonalization problem arising here at linear order in $g$, parallel closely what one would have encountered if treating quantum relativistic interacting fields in Anti-de Sitter spacetime (a brief summary can be found in \cite{madagascar}). This is not a coincidence, since nonrelativistic bosonic fields in harmonic potentials arise systematically through taking nonrelativistic limits of field systems in Anti-de Sitter spacetime \cite{BEL,BEF}. (Energy levels of quantum interacting fields in Anti-de Sitter spacetime have recently been considered from a different perspective in \cite{BSS}.)

Because the energy carried by $| \{\eta\}\rangle$ and $| \{\eta'\}\rangle$ in (\ref{blockdef}) is the same, the two annihilation operators in $\mathcal{H}_{int}$ must remove the same total amount of energy as what the two creation operators add, i.e., only terms with
 $n_1 + n_2 = n_3 + n_4$ in (\ref{eq: interaction Hamiltonian}) may contribute in the matrix elements (\ref{blockdef}). This results in a simplification that is straightforward, but has welcome analytic consequences. Namely, for the purposes of computing the shifted energy levels (\ref{enshifts}), one may replace $\mathcal{H}_{int}$ in (\ref{blockdef}) by $\mathcal{H}_{res}$ defined by
\beq
\mathcal{H}_{res}=
{\textstyle\frac12} \, \sum_{\mathclap{\substack{n_1+n_2 = n_3+n_4 \\m_1+m_2 = m_3+ m_4}}}  C_{n_1n_2n_3n_4}^{m_1m_2m_3m_4}  {\alpha}^{\dagger}_{n_1m_1}{\alpha}^{\dagger}_{n_2m_2} {\alpha}_{n_3m_3}{\alpha}_{n_4m_4}.
\label{resH}
\eeq
We note that the classical system corresponding to (\ref{resH}) is described by the Hamiltonian
\beq
\mathcal{H}_{res}=
{\textstyle\frac12} \, \sum_{\mathclap{\substack{n_1+n_2 = n_3+n_4 \\m_1+m_2 = m_3+ m_4}}}  C_{n_1n_2n_3n_4}^{m_1m_2m_3m_4}  {\alpha}^*_{n_1m_1}{\alpha}^*_{n_2m_2} {\alpha}_{n_3m_3}{\alpha}_{n_4m_4}
\label{resHclss}
\eeq
for complex-valued dynamical variables $\al_{nm}(t)$ and $\al^*_{nm}(t)$ with the symplectic form $i\sum_{nm} d\alpha^{*}_{nm} \wedge d\alpha_{nm}$. The corresponding equations of motion are
\begin{equation}
i \frac{d\alpha_{nm}}{dt} = \sum_{\mathclap{\substack{n+n_1 = n_2+n_3 \\ m+m_1 = m_2+ m_3}}} C_{nn_1n_2n_3}^{mm_1m_2m_3}  {\alpha}^{*}_{n_1m_1}{\alpha}^{}_{n_2m_2} {\alpha}_{n_3m_3}.
\label{eq: eom alphas}
\end{equation}
This classical Hamiltonian has been studied in the literature \cite{continuous,BBCE2} as the resonant approximation to the Gross-Pitaevskii equation, which is the classical limit of (\ref{Horg}) -- see also \cite{FGH} where the same resonant Hamiltonian emerges from approximating a different related PDE. Similar resonant Hamiltonian systems emerge as approximations to other physically motivated PDEs, see \cite{BEL,BEF,CF,BHP,FPU,CEV,BMR}. It is not surprising that a close relation exists between applying the resonant approximation in a classical theory and the first order of the Rayleigh-Schr\"odinger perturbation theory for the quantum version of the same problem, since both approaches succeed in approximating the time evolution of the perturbed system on time scales of order $1/g$ at small values of the coupling parameter $g$.

Studies of the classical resonant system (\ref{resHclss}-\ref{eq: eom alphas}) have produced a large set of conserved quantities (a list can be found in \cite{continuous}). Since the conserved quantities are bilinear in $\al_{nm}$ and $\al^*_{nm}$, their quantization is straightforward and does not incur ordering ambiguities. This results in a set of operators commuting with the quantum Hamiltonian (\ref{resH}) that can be presented in terms of the following combinations
\begin{align} 
{N} & = \sum_{nm}  {\alpha}^{\dagger}_{nm}{\alpha}_{nm},\label{GP2_N}\\
{E} & = \sum_{nm} n\, {\alpha}^{\dagger}_{nm}{\alpha}_{nm},\\
{M} & = \sum_{nm} m\, {\alpha}^{\dagger}_{nm}{\alpha}_{nm},\\
{Z}_+ &=  \sum_{nm}  \sqrt{\frac{n+m+2}{2}} \, {\alpha}^{\dagger}_{n+1,m+1}{\alpha}_{nm},\label{GP2_Zp}\\
{Z}_- &=  \sum_{nm} \sqrt{\frac{n-m+2}{2}} \, {\alpha}^{\dagger}_{n+1,m-1}{\alpha}_{nm},\\
{W} &= \sum_{nm}  \frac{\sqrt{n^2-m^2}}{2} \, {\alpha}^{\dagger}_{nm}{\alpha}_{n-2,m}.\label{GP2_W}
\end{align}
The last three operators are not Hermitian, and their Hermitian conjugates $Z_+^\dagger$, $Z_-^\dagger$ and $W^\dagger$ should also be included. We do not explicitly give the summation ranges of $n$ and $m$, but it is understood, here and elsewhere, that summations run over all possible values of $n$ and $m$ for which the creation-annihilation operators in the summand correspond to existing modes. As a reminder, $\alpha_{nm}$ corresponds to an actual oscillator mode if $n\ge 0$, $|m|\le n$, and $n-m$ is even.

While the conservation of $N$, $E$ and $M$ is straightforwardly seen from the general structure of the resonant Hamiltonian (\ref{resH}), the conservation of $Z_+$, $Z_-$ and $W$ relies on the specific form of the interaction coefficients $C$ given by (\ref{intcoeff}) and is not immediately obvious. These conservation laws were established in \cite{FGH} where the resonant system (\ref{resHclss}-\ref{eq: eom alphas}) was derived for bosons with contact interactions {\it without} a harmonic trap. The reason the same resonant system is of relevance both with and without a harmonic trap is that, specifically in two dimensions and for contact interactions, the two situations are mapped into each other using the so-called `lens transform,' also known as the pseudo-conformal compactification \cite{Carles,Tao}. Thus, as explicitly pointed out in \cite{continuous}, the same conservation laws are respected in the presence of a harmonic trap. As the original derivations of the conservation laws in the mathematical literature would have taken us rather far outside our present context, we feel it beneficial to present elementary proofs, which we give in Appendix \ref{App_A}. The essence of these proofs is that the harmonic oscillator mode functions (\ref{psipossgn}) are expressed through the Laguerre polynomials, and identities for the Laguerre polynomials imply that the interaction coefficients defined by (\ref{intcoeff}) satisfy certain finite difference equations with respect to their mode number indices. These finite difference equations, in turn, imply the conservation of $Z_+$, $Z_-$ and $W$.

\begin{figure}
\centering
\includegraphics[scale=0.3]{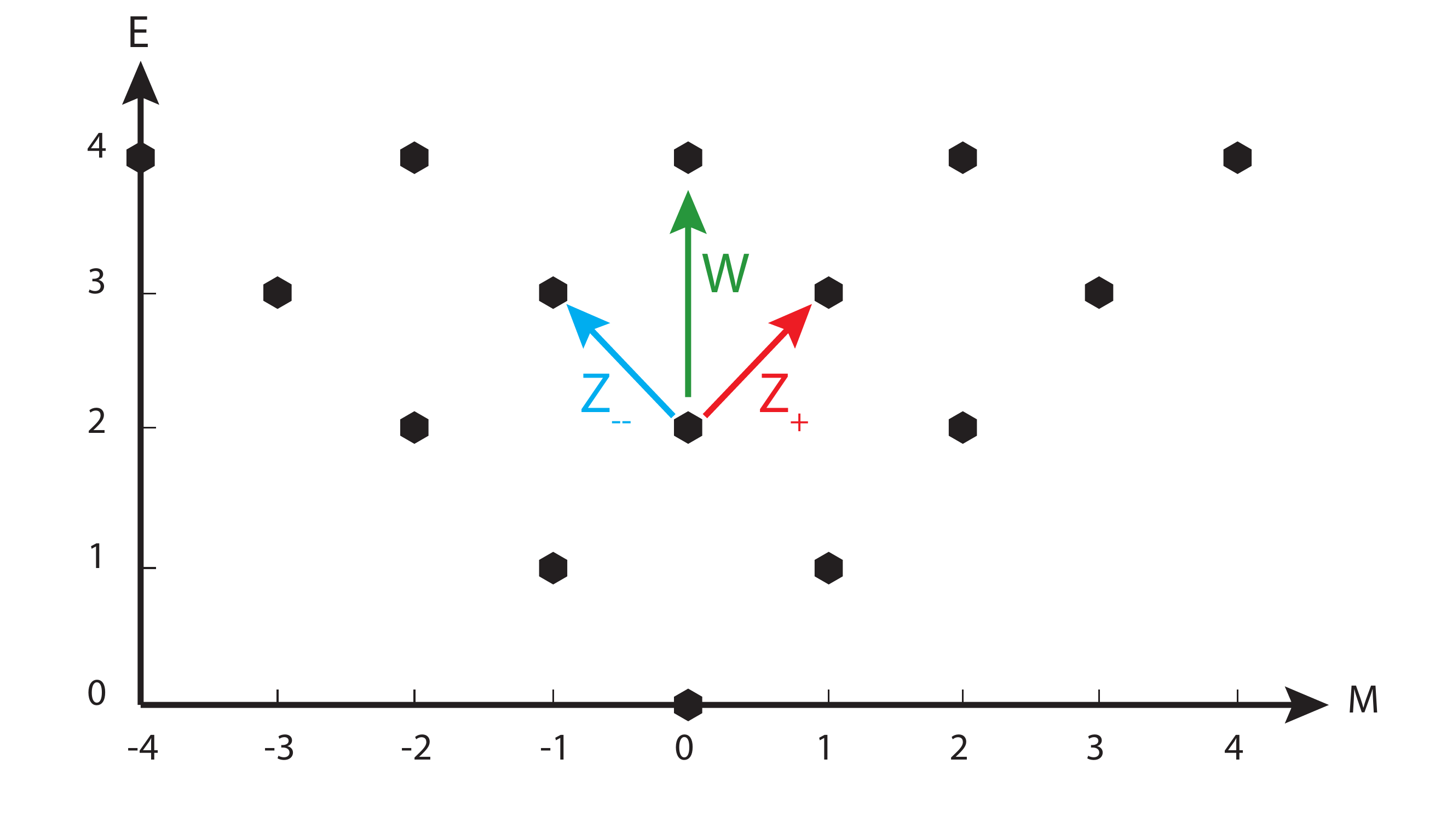}
\caption{The action of raising operators $Z_+$, $Z_-$ and $W$ in the $(M,E)$-plane. The patterns of consecutive moves generated by $Z_+$, $Z_-$ and $W$ are closely reminiscent of (though not exactly identical to) the so-called Motzkin walks. (Definitions and some recent applications to quantum spin chains can be found in \cite{motzkin1,motzkin2}.)}
\label{GP2_raising}
\end{figure}

What do the conserved quantities (\ref{GP2_N}-\ref{GP2_W}) tell us about the diagonalization of (\ref{resH}), and consequently about the energy shifts (\ref{enshifts})? First of all, since $N$, $E$ and $M$ commute not only with $\mathcal{H}_{res}$ but also with each other, the four operators can be diagonalized simultaneously, grouping the eigenvalues of $\mathcal{H}_{res}$ into $(N, E, M)$-blocks, labelled by the integer eigenvalues of $N$, $E$ and $M$ (which equal $\sum\eta_{nm}$, $\sum n\,\eta_{nm}$ and $\sum m\,\eta_{nm}$, respectively). In fact, since $N$, $E$ and $M$ are diagonal in the Fock basis, the block-diagonal structure of $\mathcal{H}_{res}$ is seen directly by considering its Fock basis matrix elements. Evidently, there are finitely many states for a given triplet $(N,E,M)$. Each such block describes energy shifts (\ref{enshifts}) of the unperturbed level of energy $E$ in the sector with $N$ particles carrying a total of $M$ units of angular momentum. Now, the action of $Z_+$ moves any state in an $(N,E,M)$-block to a state in the $(N,E+1,M+1)$-block, the action of $Z_-$ moves any state in an $(N,E,M)$-block to a state in the $(N,E+1,M-1)$-block, and the action of $W$ moves any state in an $(N,E,M)$-block to a state in the $(N,E+2,M)$-block, as depicted in Fig.~\ref{GP2_raising}.

Since  $Z_+$, $Z_-$ and $W$ commute with $\mathcal{H}_{res}$, their action will copy the eigenvalues from lower to higher $(N,E,M)$-blocks. A distinctive signature of this structure is that many energy shifts $\eps_I$ in (\ref{enshifts}) will be identical for different unperturbed energy levels, and thus infinitely many integer energy differences of the noninteracting boson spectrum will survive at linear order in $g$. This is in fact a reflection of the general pattern valid for any $g$, translated into the resonant system (\ref{resH}). Indeed, the center-of-mass motion is known to separate for bosons with arbitrary translation-invariant interactions confined in a harmonic trap (see, e.g., \cite{bbbb}). The energy of the center-of-mass, given by an energy eigenvalue of a two-dimensional harmonic oscillator, is simply added to the energy of the relative motion. The two-dimensional harmonic oscillator admits two independent raising operators increasing the energy value by 1 unit, which corresponds in the language of the resonant system to the action of $Z_+$ and $Z_-$. Additional symmetry enhancement exists specifically for bosons with contact interactions in two spatial dimensions, where the symmetry is extended to the full Schr\"odinger group \cite{Niederer,OFN}. This results in the existence of the Pitaevskii-Rosch breathing mode and the corresponding independent raising operator increasing the energy by 2 units \cite{breathing}. This corresponds to the action of $W$. We note that the raising operators of the original system are not symmetries in the standard Hamiltonian sense (they have nontrivial commutation relations with the Hamiltonian), but they are translated into ordinary Hamiltonian symmetries (operators commuting with the Hamiltonian) in the language of the resonant system (\ref{resH}).

While the symmetries of the resonant system (\ref{resH}) originate from the known symmetry structures of the original problem (\ref{Horg}), their operation has more powerful consequences. Indeed, under normal circumstances, symmetries allow generating new energy eigenstates from a known energy eigenstate, but do not simplify the process of finding additional energy states outside a given symmetry multiplet. In our case, the spectrum of (\ref{resH}) splits into finite-sized blocks (originating from the degenerate energy levels of noninteracting bosons), which allows for computing parts of the spectrum exactly, independently of other parts of the spectrum. Then, for example, instead of computing all eigenvalues in blocks up to a certain energy, one can simply diagonalize a single sufficiently large block, and then recover the lower blocks by repeatedly acting with $Z_\pm^\dagger$ and $W^\dagger$ on the explicitly found eigenvectors. (Alternatively, one can diagonalize operators like $Z_+Z_+^\dagger$ simultaneously with $\mathcal{H}_{res}$, and this will allow for recovering energy eigenvalues from lower blocks, as we shall briefly describe in section \ref{generic}.) Separation of energy eigenstates with respect to the action of the symmetries (\ref{GP2_Zp}-\ref{GP2_W}) will also play a crucial role in our analysis of the level spacing statistics.
Furthermore, a number of mathematical results are known for the classical counterpart (\ref{resHclss}) of our system (\ref{resH}), such as bounds on the classical Hamiltonian, and they have implications for the quantum problem we are considering.
Our goal for the rest of this treatment will be to systematically explore these issues, and to comment on the eigenvalue statistics in hope of addressing the question whether the set of symmetries we have used is complete. As considerations for general $(N,E,M)$-blocks become rather involved, we shall start by focusing on blocks with $E=M$, known as the Lowest Landau Level (LLL), where the algebra simplifies considerably, and all essential concepts and techniques may be more transparently demonstrated. We shall thereafter proceed with our analysis of general $(N,E,M)$-blocks.


\section{The LLL sector}\label{section_LLL}

\subsection{Classical and quantum LLL truncation}

It is well-known \cite{continuous, GT, BBCE, GGT} that the classical resonant system (\ref{resHclss}-\ref{eq: eom alphas}) can be consistently truncated to the set of modes satisfying $n=m$, the so-called Lowest Landau Level (LLL) sector. Setting all modes with $n\ne m$ to zero in the initial state guarantees that they never get excited by the evolution equation (\ref{eq: eom alphas}). (The same set of LLL modes has often appeared in more phenomenological studies of rapidly rotating trapped Bose-Einstein condensates \cite{fetter}.)

Classical truncations of this sort in general have no direct implications for the quantum theory, since the uncertainty principle makes it impossible to set canonical variables to zero. The LLL truncation happens to have a direct translation to the quantum system (\ref{resH}), nonetheless, for reasons that are essentially kinematical. Consider an $(N,E,M)$-block of Fock states with $N$ particles, $E$ units of energy and $M$ units of angular momentum, and impose $E=M$. Then, the occupation numbers $\eta_{nm}$ in the corresponding Fock states must vanish for $n\ne m$. Indeed, only modes with $m\le n$ exist, and having any modes with $m<n$ excited guarantees that the total angular momentum $M$ is less than the total energy $E$, in contradiction with our assumption. Thus, any block with $E=M$ is entirely composed of LLL states. Creation-annihilation operators corresponding to non-LLL modes do not contribute to matrix elements (\ref{blockdef}) between any two states in an $E=M$ block. We shall refer to the $E=M$ blocks as `LLL blocks' for obvious reasons.

In view of the above picture, for the analysis of level splitting in the LLL blocks, the full resonant Hamiltonian (\ref{resH}) can be replaced by the following simpler LLL Hamiltonian
\begin{align} \label{resonant_LLL}
\mathcal{H}_{LLL} = {\textstyle\frac{1}{2}} \hspace{-5mm}\sum^{\infty}_{\substack{n_1,n_2,n_3,n_4=0\\n_1 +n_2= n_3+ n_4}}\hspace{-5mm}C_{n_1n_2n_3n_4} \ad_{n_1} \ad_{n_2} \alpha_{n_3} \alpha_{n_4}.
\end{align}
Here, we have relabelled $\al_{n_im_i}$ with $n_i=m_i$ as simply $\al_{n_i}$.
The LLL interaction coefficients can be directly evaluated \cite{continuous} as a special case of (\ref{intcoeff}) resulting in the following simple expression:
\begin{align}
C_{n_1n_2n_3n_4} = \frac{((n_1+n_2+n_3+n_4)/2)!}{2^{n_1+n_2}\sqrt{n_1!n_2!n_3!n_4!}}.
\label{CLLL}
\end{align}

The classical system corresponding to the LLL Hamiltonian (\ref{resonant_LLL}) has been studied as an approximation to the Gross-Pitaevskii equation for Bose-Einstein condensates \cite{continuous, GT,BBCE,GGT}, and possesses many special properties, for example it admits an invariant manifold where the nonlinear equations can be solved exactly, and the solutions show interesting long-term return behaviors \cite{BBCE}. Spatial positions of zeros of the wavefunctions (known as `vortices') corresponding to some solutions may also be analyzed to a great extent \cite{GGT}. This system is a representative of a very large class of partially solvable resonant systems developed in \cite{AO}, which are of the form (\ref{resonant_LLL}) but with different choices of the interaction coefficients $C$, and which share many of the special dynamical features we have just mentioned.

The quantum Hamiltonian (\ref{resonant_LLL}-\ref{CLLL}), which is what is of interest for us here, inherits the following conserved quantities from (\ref{GP2_N}-\ref{GP2_W}):
\begin{align}
&{N} = \sum_{k=0}^\infty {\alpha}^\dagger_k {\alpha}_k, \qquad {E} = \sum_{k=1}^\infty k {\alpha}^\dagger_k {\alpha}_k,\nonumber\\
 &{Z} = \sum_{k=0}^\infty \sqrt{k+1} \,\alpha^\dagger_{k+1} \alpha_k.\label{ZLLL}
\end{align}
The commutators of these operators with $\mathcal{H}_{LLL}$ vanish, and the commutators among themselves and with $Z^\dagger$ vanish except for
\begin{align}
& [{E}, {Z}] = {Z}, \quad [{E}, {Z}^\dagger]=-{Z}^\dagger,\nonumber\\
& [Z,Z^\dagger]=-N.\label{NEZcomm}
\end{align}

Energy spectra of systems of the form (\ref{resonant_LLL}) with general interaction coefficients $C$ have been studied numerically in \cite{quantres}. We shall now examine how the specific symmetry structures emerging for $C$ given by (\ref{CLLL}) influence the eigenvalue patterns.

\subsection{Structure of the Hamiltonian blocks}\label{strspc}

The structure of the eigenvalue problem for the LLL Hamiltonian \eqref{resonant_LLL} is inherited as a simplified version from what has been described in the previous section for the full resonant Hamiltonian (\ref{resH}). On the other hand, the generalities of diagonalizing Hamiltonians of the form (\ref{resonant_LLL}) with arbitrary interaction coefficients $C$ have been spelled out in \cite{quantres}, and we shall closely follow that treatment. 

One starts with defining the LLL Fock basis as
\beq
|\eta_0,\eta_1,\dots\rangle=\prod_{k=0}^\infty\frac{(\ad_k)^{\eta_k}}{\sqrt{\eta_k!}}|0,0,0,\dots\rangle,
\eeq
such that
\begin{align}
{\alpha}_k^\dagger {\alpha}_k |\eta_0,\eta_1,\dots\rangle= \eta_k |\eta_0,\eta_1,\dots\rangle 
\end{align}
for any $k$, and $\eta_k$ are nonnegative integers.
The eigenvalues of $N$ and $E$ in this basis are evidently
\begin{align}
N = \eta_0 + \sum_{k=1}^\infty \eta_k, \quad E = \sum_{k=1}^\infty k \, \eta_k.
\end{align}
The Hamiltonian \eqref{resonant_LLL} has nonvanishing matrix elements between $ |\eta_0,\eta_1,\dots\rangle$ and $ |\eta'_0,\eta'_1,\dots\rangle$ only for two sets of occupation numbers $\lbrace \eta_k \rbrace$ and $\lbrace \eta'_k \rbrace$ having the same values of $N$ and $E$. Thus, the Hamiltonian is block-diagonal in the Fock basis, where the blocks are labeled by the nonnegative numbers $(N,E)$. The number of states in an $(N,E)$-block is given by the number of integer partitions of $E$ into at most $N$ parts  \cite{quantres}, a well known number-theoretic function usually denoted as $p_N(E)$. One thus has to diagonalize finite-sized $p_N(E)\times p_N(E)$ matrices to get the eigenvalues of (\ref{resonant_LLL}) within each $(N,E)$-block.

Up to this point, our discussion of the diagonalization has been generic and did not make any reference to the specific form of the interaction coefficients given by (\ref{CLLL}). For this specific form of the interaction coefficients, an extra conserved operator $Z$ given by (\ref{ZLLL}) and its Hermitian conjugate enter the game.
These operators act as raising and lowering operators for $E$: namely, acting on a state in an $(N,E)$-block, $Z$ produces a state in the $(N,E+1)$-block, and $Z^\dagger$, a state in the $(N,E-1)$-block. Since $Z$ and $Z^\dagger$ commute with $\mathcal{H}_{LLL}$, eigenvectors of the latter are mapped into eigenvectors by this action, while their eigenvalues remain intact.

We are thus brought to the first qualitative conclusion of our analysis. By the action of $Z$, eigenvalues are recursively copied from lower $(N,E)$-blocks to higher $(N,E)$-blocks. Thus, each eigenvalue is present in infinitely many copies. For any two blocks $(N,E_1)$ and $(N,E_2)$ with $E_1<E_2$, the  $\mathcal{H}_{LLL}$ eigenvalues of the first block are a subset of the eigenvalues of the second block. The energy shifts $\eps_I$ in (\ref{enshifts}) for the two corresponding unperturbed levels are evidently in the same relation.

How do new eigenvalues emerge in this picture as we move to higher values of $E$? The action of $Z$ copies all the $p_N(E)$ eigenvalues of the $(N,E)$-block into the $(N,E+1)$-block. The latter block has $p_N(E+1)>p_N(E)$ eigenvalues, and the excess eigenvalues must correspond to eigenvectors annihilated by $Z^\dagger$. Indeed, if they were not annihilated by $Z^\dagger$, the action of $Z^\dagger$ would have produced an eigenvector the $(N,E)$-block with the same eigenvalue, in contradiction with our assumption that the eigenvalue is new. Thus, any new eigenvalues in the $(N,E+1)$-block compared to the $(N,E)$-block must come from vectors $|\Psi\rangle$ in the kernel of $Z^\dagger$, 
\beq
Z^\dagger|\Psi\rangle =0.
\label{LLLker}
\eeq
In order for the whole picture to be consistent, the dimension of this kernel within the $(N,E+1)$-block must precisely account for the difference in the dimensions of the $(N,E)$- and  $(N,E+1)$-blocks:
\begin{align}
\text{\rm dim}_{\sssty (N,E+1)} \left( \ker {Z}^\dagger \right) = p_N(E+1)-p_N(E).
\label{kerdim}
\end{align}
To see this structure more explicitly, note first that $Z$ cannot annihilate a nonvacuum state. Indeed, by (\ref{NEZcomm}),
\beq
\big|Z|\Psi\rangle\big|^2=\big|Z^\dagger|\Psi\rangle\big|^2+\langle\Psi|N|\Psi\rangle\ge\langle\Psi|N|\Psi\rangle.
\eeq
Thus, by acting with $Z$ on a complete basis $|\Psi_I\rangle$ in the $(N,E)$-block containing $p_N(E)$ elements, we obtain $p_N(E)$ linearly independent vectors $Z|\Psi_I\rangle$ in the $(N,E+1)$-block. Consider the orthogonal complement of the subspace spanned by these vectors within the $(N,E+1)$-block, which by construction has $ p_N(E+1)-p_N(E)$ dimensions. Any of the vectors $|\Phi\rangle$ in this orthogonal complement must be annihilated by $Z^\dagger$ since, on the one hand,  $Z^\dagger|\Phi\rangle$  belongs to the $(N,E)$-block, and on the other hand it is orthogonal to all states in the $(N,E)$-block, since $\langle \Psi_I|Z^\dagger|\Phi\rangle=0$ as a consequence of $|\Phi\rangle$ being orthogonal to $Z|\Psi_I\rangle$. Therefore any such $|\Phi\rangle$ is in the kernel of $Z^\dagger$, while none of the vectors $Z|\Psi_I\rangle$ can be in the kernel of $Z^\dagger$ as $|Z^\dagger Z|\Psi_I\rangle|^2>0$ by (\ref{NEZcomm}). We have thereby established (\ref{kerdim}).

The picture we have presented leads to simplifications in generating the spectrum of $\mathcal{H}_{LLL}$ numerically. Instead of individually diagonalizing all the $(N,E)$-blocks, as one does for generic resonant systems \cite{quantres}, we may simply choose one block with a sufficiently large $E$, and diagonalize explicitly only this block. All the eigenvalues of the blocks with the same $N$ and lower $E$ are contained in this block due to the inheritance property we have described. These eigenvalues of the lower blocks can be extracted by either repeatedly acting on the eigenvectors of the given $(N,E)$-block with $Z^\dagger$, or by diagonalizing $ZZ^\dagger$ simultaneously with $\mathcal{H}_{LLL}$ in the given $(N,E)$-block. Indeed, the eigenvectors in the kernel (\ref{LLLker}) will be annihilated by $ZZ^\dagger$, while nonzero eigenvalues of $ZZ^\dagger$ will mark the energy eigenstates that have descended from the lower values of $E$ via the action of $Z$. The eigenvalues of $ZZ^\dagger$ can be obtained from the commutation relations (\ref{NEZcomm}), and they depend on the value of $E$ in the block where a given energy eigenvalue first appears before being transported to the current block by the action of $Z$. This is why inspecting the eigenvalues of $ZZ^\dagger$ allows for a recovery of the lower $(N,E)$-blocks from a given block.

One particular eigenvalue stands out prominently in the inheritance process we have described. One can start with the $(N,0)$-block that contains only one state, $|N,0,0,\ldots\rangle$, which is of course an eigenstate of $\mathcal{H}_{LLL}$,
\beq
\mathcal{H}_{LLL}|N,0,0,\ldots\rangle=\frac{N(N-1)}2 |N,0,0,\ldots\rangle.
\label{N000}
\eeq
The corresponding eigenvalue $N(N-1)/2$ will be propagated by the action of $Z$ to all the higher $(N,E)$-blocks, where it will correspond to the vector $Z^E|N,0,0,\ldots\rangle$. We shall now derive bounds on $\mathcal{H}_{LLL}$, which make it clear that this eigenvalue is always the highest one in the spectrum, while the other eigenvalues of an $(N,E)$-blocks lie between $0$ and $N(N-1)/2$.

\subsection{Bounds on the LLL Hamiltonian}\label{boundLLL}

In establishing bounds on $\mathcal{H}_{LLL}$, the following explicit representation of the resonant summation in (\ref{resonant_LLL}) will prove convenient:
\beq
\mathcal{H}_{LLL}= \frac{1}{2} \sum^{\infty}_{j=0} \sum^{j}_{n,k=0} C^{LLL}_{n,j-n,k,j-k} \alpha_{n}^{\dagger} \alpha_{j-n}^{\dagger} \alpha_{k} \alpha_{j-k}.
\label{eq: LLL hamiltonian}
\eeq
(Resonant summations in this form have been extensively employed in \cite{meloturb}.) From this representation and (\ref{CLLL}), one may immediately write
\begin{equation}
\mathcal{H}_{LLL} = \frac{1}{2} \sum^{\infty}_{j=0} \frac{j!}{2^{j}} A_j^{\dagger} A_j,
\end{equation}
where
\beq
A_j= \sum^{j}_{n=0} \frac{\alpha_{n} \alpha_{j-n}}{\sqrt{n!(j-n)!}}.
\eeq
Thus, $\mathcal{H}_{LLL}$ is manifestly a nonnegative operator and all of its eigenvalues are nonnegative as claimed under (\ref{N000}).

We then proceed to establish the upper bound on $\mathcal{H}_{LLL}$, which is slightly less straightforward. Consider
\beq
\mathcal{H}_{inv} = \frac{{N}({N}-1)}{2} - \mathcal{H}_{LLL}.
\eeq
Similarly to the previous proof, we wish to write $\mathcal{H}_{inv}$ as a sum of manifestly nonnegative terms. Using $[\alpha^\dagger_l,\alpha_k]=-\delta_{kl}$, we first note that
\begin{align}
\frac{{N}({N}-1)}{2}&=\frac12\sum_{k,l=0}^\infty \ad_k\ad_l\al_k\al_l\\
&=\frac12\sum_{j=0}^\infty\sum_{k=0}^j \ad_k\ad_{j-k}\al_k\al_{j-k}.\nonumber
\end{align}
We then insert into this representation of $N(N-1)/2$ a tautological decomposition of unity in terms of the binomial identity
\beq
1=\frac1{2^j}\sum_{l=0}^j\frac{j!}{l!(j-l)!}.
\label{binomid}
\eeq
With these preliminaries, one can straightforwardly write
\begin{equation}
\mathcal{H}_{inv} =\frac14 \sum^{\infty}_{j = 0} \frac{j!}{2^j}\sum_{k,l = 0}^{j} A_{kl}^{j\dagger} A^j_{kl},
\label{eq: Htilde}
\end{equation}
where
\begin{align}\label{eq: Htilde_A}
A_{kl}^j =\frac{\al_k \, \al_{j-k}}{\sqrt{l!(j-l)!}} - \frac{\al_l \, \al_{j-l}}{\sqrt{k!(j-k)!}}.
\end{align}
This proves that $\mathcal{H}_{LLL}$ is bounded from above by $N(N-1)/2$. 

To summarize, all the eigenvalues $\mathcal{H}_{LLL}$ within an $(N,E)$ block lie between 0 and the largest eigenvalue $N(N-1)/2$, which corresponds to the state $Z^E|N,0,0,\ldots\rangle$ as we have previously claimed. 

\subsection{Level spacing statistics}\label{levelsp}

We have seen in the above treatment that the known symmetries of the LLL Hamiltonian have powerful implications for the structure of its spectrum. A question naturally arises whether the symmetries generated by (\ref{ZLLL}) are all there is, or there are extra conserved operators in addition to $N$, $E$ and $Z$. While there is no algorithmic way to construct extra conserved operators or prove their absence, quantum chaos theory \cite{GMW,haake} provides an attractive set of tools to shed light on this question. The bulk of quantum chaos theory revolves around a set of conjectures derived from extensive numerical experimentation, and we will thus not have water-tight theorems at our disposal. Nonetheless, we feel that the indicators supplied by this approach are of great practical use in our context.

A key tenet of quantum chaos theory is that the quantum spectra of classically chaotic systems have distances between neighboring energy levels that obey a qualitatively different statistics from the quantum spectra of classically integrable systems. More specifically, distances between energy eigenvalues for a generic integrable system follow \cite{btint} the Poisson distribution
\beq
\rho_{Poisson}(s)=e^{-s}.
\label{pois}
\eeq
Here, $s$ are distances between neighboring levels of a properly normalized (more technically, `unfolded') energy level sequence that we shall discuss below. The distribution (\ref{pois}) is the  same as the distribution of distances between points randomly thrown on a line with a unit mean density. For a chaotic system, on the other hand, one expects \cite{BGS} the Wigner-Dyson distribution, which is well approximated for practical purposes by the `Wigner surmise'
\beq
\rho_{Wigner}(s)=\frac{\pi s}2\, e^{-\pi s^2/4}.
\label{wign}
\eeq
This is the statistics of distances between eigenvalues of real symmetric random matrices with independent identically distributed entries. Note that (\ref{wign}) vanishes at zero energy level separation, a phenomenon known as `level repulsion.'

We have to clarify what is meant by the `unfolding' procedure used to generate the normalized level distances $s$ from a given set of energy levels. The need for unfolding is most easily understood in the original context of the quantum chaos theory, that is, the energy spectra of chaotic quantum-mechanical systems with a few degrees of freedom, such as quantum billiards \cite{BGS}. In this situation, there are infinitely many energy levels and the distances between the neighboring levels may become progressively longer or shorter as one moves to higher energies. Thus, one is not even guaranteed to have the distribution of plain energy distances converge to a limit as more and more energy levels are included. What one must do is differentially rescale the spectrum so that on intervals containing many energy levels the mean density is 1 in any part of the spectrum. Then, the common distribution function for level distances will exist, and it is the level distance distributions in thus unfolded level sequences for which the quantum chaos conjectures are formulated. A contemporary discussion of unfolding can be found in \cite{GMW,abdulmagd,luukko}. 

Unfolding may seem more subtle for inherently finite samples of energy levels, such as the levels within the $(N,E)$-blocks of the LLL Hamiltonian. Nonetheless, the following simple-minded definition \cite{quantres} will prove perfectly effective for our purposes. Consider a sample with $K$ eigenvalues denoted $E_I$ with $I=1,\ldots,K$. We shall assume $K$ to be reasonably large (comparable to 1000 in our numerics). One can choose an integer $\Delta$ close to $\sqrt{K}$, and differentially stretch the spectrum so that the level density is 1 on intervals containing $2\Delta$ energy levels in any part of the spectrum. To this end, define the raw unfolded sequence of level spacings
\beq
s_I^{(raw)}=\frac{E_{I+1}-E_{I}}{E_{I+\Delta}-E_{I-\Delta}},
\eeq
with $I=\Delta+1,\Delta+2,\ldots,K-\Delta$. The division by $(E_{I+\Delta}-E_{I-\Delta})$ factors out the mean level density over intervals containing  $2\Delta$ adjacent levels and centered on the point of observation. We then compute the average of this raw sequence $\bar s=(\sum_{I=\Delta+1}^{K-\Delta}s_I^{(raw)})/(K-2\Delta)$, and define the normalized unfolded sequence
\beq
s_I=\frac{s_I^{(raw)}}{\bar s},
\eeq
whose mean by construction equals 1.
This is our definition of $s$ for comparing with (\ref{pois}) and (\ref{wign}). While the smoothing parameter $\Delta$ and the bin size for plotting distribution histograms have to be chosen by hand (though $\sqrt{K}$ is the natural scale for both), there are no free parameters in the distributions (\ref{pois}) and (\ref{wign}) and thus no fitting involved in the comparison after the histograms have been plotted.

We note that the chaotic Wigner-Dyson distribution, which we have approximately represented by the Wigner surmise (\ref{wign}), is only expected to emerge if the spectrum is separated into blocks according to the values of all pairwise commuting conserved operators (the analogs of globally defined classical integrals of motion in involution). The underlying philosophy is that for each specific value of such commuting conserved operators, one essentially has an independent Hamiltonian system, and the spectra of these individual Hamiltonian systems are not expected to be correlated. Then, lumping together the different blocks for computing the level spacing statistics would wash out the level repulsion phenomenon, which is responsible for the shape of (\ref{wign}) and requires correlations between the energy levels (see for example section 3.1 of \cite{DKPR}).

This is precisely what one could see in the numerical analysis of the LLL Hamiltonian in \cite{quantres}. In that analysis, guided by the general structure of the eigenvalue problems for resonant Hamiltonians, the eigenvalues were split into $(N,E)$-blocks. Within such blocks, a Poissonian distribution of unfolded level spacings was seen, and the question was raised about its implications. We now answer this question by noticing that $N$ and $E$ are not the only quantities  commuting with each other and $\mathcal{H}_{LLL}$, as $ZZ^\dagger$ commutes with all the three. One must then separate the LLL Hamiltonian eigenvalues into blocks not only according to their values of $N$ and $E$, but also according to the values of $ZZ^\dagger$.

In practice, because of the structure of the eigenvalue problem of $\mathcal{H}_{LLL}$ described in section \ref{strspc}, it is sufficient to consider the eigenvectors of  $\mathcal{H}_{LLL}$ annihilated by $ZZ^\dagger$, i.e., the eigenvectors in the kernel of $Z^\dagger$. Indeed, all other $ZZ^\dagger$-blocks are simply inherited through repeated action of $Z$ on the $Z^\dagger$-kernel of some other $(N,E)$-block with a lower value of $E$. So the variety of statistics present in the problem is fully exhausted by looking at $Z^\dagger$-kernels only.

\begin{figure}[t]
\includegraphics[scale=0.3]{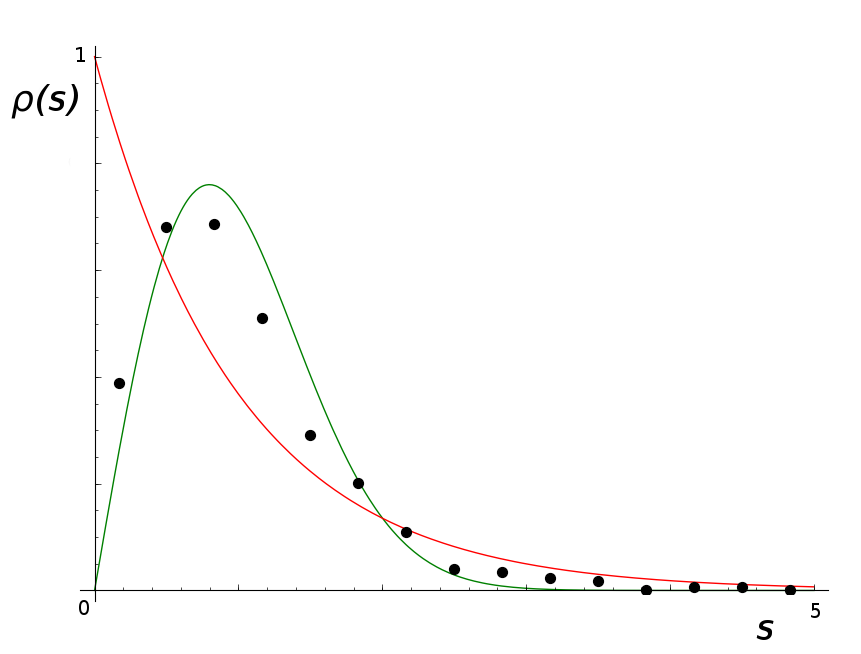}
\caption{The Wignerian distribution of the unfolded level spacings in the $Z^\dagger$-kernel of the $(27,27)$-block of the LLL Hamiltonian. The green curve is the Wigner surmise (\ref{wign}), while the red curve is the Poisson distribution (\ref{pois}).}
\label{LLL2727}
\end{figure}

We have implemented this procedure for the $N=E=27$ block that has already been considered in \cite{quantres}. If the joint level spacing statistics in the entire block is plotted, one sees a Poissonian distribution, as in \cite{quantres}. If the levels outside the $Z^\dagger$-kernel are excised, one obtains the distribution in Fig.~\ref{LLL2727}. This distribution is very far from the Poissonian shape, and close to the Wigner-Dyson shape, which is a strong argument against integrability. (For an integrable system, one expects the distribution to be Poissonian in all large blocks, along the lines of the Berry-Tabor results \cite{btint}.) One may notice that the actual distribution in Fig.~\ref{LLL2727} is slightly shifted to the left compared to the Wigner-Dyson curve. This shift, in fact, increases when one moves to other blocks away from $N=M$. There, the level repulsion decreases, and one gets what is known as crossover behaviors. For $N\ll M$, the distribution acquires Poissonian features. This does not change our conclusions on integrability, but creates room for some subtle structures. The actual distributions are given in Appendix~\ref{App_B}.

\subsection{Generalizations}\label{LLLgen}

We briefly comment on possible generalizations of the material of this section that take us outside the immediate scope of the present article. Resonant systems of the form (\ref{resonant_LLL}), with various assignments of the interaction coefficients $C$, have appeared in studies of weakly nonlinear problems emerging from a range of physical applications. Apart from cold atomic gases and Bose-Einstein condensates, there are applications in high-energy and gravitational physics, in particular due to connections to the topic of instability of the Anti-de Sitter spacetime \cite{BR,rev2}, and dynamics of related nonlinear wave equations in highly symmetric spacetimes \cite{CF,BEL}. Such resonant systems are also of relevance to the Hartree equation in harmonic potentials \cite{BEF}.

A number of resonant systems emerging from such weakly nonlinear considerations \cite{CF,BEL,BEF,BBCE2} possess features closely reminiscent of what we have used above to analyze the resonant LLL Hamiltonian. In \cite{AO}, these features have been distilled into a formulation of a very large class of resonant systems that, in particular, admit an analog of the $Z$-conservation that has played a crucial role in the derivations of this section. In \cite{quantres}, numerical analysis of the quantum version of these systems was performed, demonstrating a number of distinctive traits, such as the eigenvalue inheritance from lower to higher values of $E$ in the $(N,E)$-blocks, and the maximal eigenvalue within each block being equal to $N(N-1)/2$. We have seen in this section how these observations can be proved for the LLL system. Analogous proofs will hold for other systems in the large class formulated in \cite{AO}. In particular, the $Z$-conservation will directly translate into eigenvalue inheritance, while an upper bound on the eigenvalues will follow (at least for systems with positive interaction coefficients) from a generalization of the summation identity (\ref{binomid}) that plays a defining role in the constructions of \cite{AO}. This opens up space for applications of the material of this section, in a slightly modified form, to a number of very different physical problems.


\section{Generic energy levels}\label{generic}

As we turn to generic $(N,E,M)$-blocks of the resonant Hamiltonian (\ref{resH}), we essentially have to retrace the steps taken in the previous section for the LLL blocks with $E=M$, but the amount of technical effort required increases. A large part of the reason is that the interaction coefficients (\ref{intcoeff}) no longer have a simple closed form like (\ref{CLLL}) and are expressed through the Laguerre polynomials as
\begin{align}
&C^{m_1m_2m_3m_4}_{n_1n_2n_3n_4} = 2\left(\prod_{i=1}^4\sqrt{\frac{(\frac12(n_i-m_i ))!}{(\frac12(n_i+m_i ))!}}\right)\label{CLaguerre}\\ 
&\times\int^{\infty}_{0} \hspace{-1.5mm}d\rho\, \,e^{-2\rho} \rho^{(m_1+m_2+m_3+m_4)/2}\left(\prod_{i=1}^4 L^{m_i}_{\frac{n_i-m_i}{2}}(\rho)\right).\nonumber
\end{align}
There is also a much bigger set of conserved operators (\ref{GP2_N}-\ref{GP2_W}) that must be taken into account.
(Note that integrals of quadruple products of Laguerre polynomials of the form (\ref{CLaguerre}) have many special properties. For example, for the radially symmetric case $m_i=0$, there is appreciable mathematical literature on positivity \cite{lagpos} and combinatorial interpretations \cite{lagcomb1,lagcomb2} of such integrals.)

As for the LLL case, there is inheritance of eigenvalues of $\mathcal{H}_{res}$ from blocks with lower $E$ and $M$ to higher blocks via the action of $Z_\pm$ and $W$ as depicted on Fig.~\ref{GP2_raising}. New eigenvalues, not found in the lower blocks, always enter through the joint kernel of $Z_\pm^\dagger$ and $W^\dagger$, i.e., the subspace of states $|\Psi\rangle$ satisfying
\beq
Z_\pm^\dagger|\Psi\rangle=0,\qquad W^\dagger|\Psi\rangle=0.
\label{kern3}
\eeq
Similarly to the previous section, acting on the state where all the $N$ particles occupy the lowest energy mode with different arrangements of $Z_\pm$ and $W$ results in states of energy $N(N-1)/2$ in each of the higher $(N,E,M)$-blocks. Unlike the LLL sector, these states may have high multiplicities within their $(N,E,M)$-blocks as there are many inequivalent ways to reach the same $(N,E,M)$-block acting with different arrangements of $Z_\pm$ and $W$.
As for the LLL sector, $N(N-1)/2$ is the highest energy state within each $(N,E,M)$-block as a consequence of bounds on the resonant Hamiltonian (\ref{resH}).

Establishing bounds on  $\mathcal{H}_{res}$ is considerably more involved than for the case of  $\mathcal{H}_{LLL}$, and we give the proofs in Appendix~\ref{App_C}. These proofs retrace the mathematical literature \cite{FGH,continuous}, but in a language more accessible to physicists.
The result is that all the eigenvalues in general  $(N,E,M)$-blocks lie between $0$ and the maximal eigenvalue $N(N-1)/2$, as in the previous section.

The conserved operators $N$, $E$, $M$,  $Z_\pm$, $Z_\pm^\dagger$, $W$ and $W^\dagger$ obey the following commutator algebra, where we only give the nonzero commutators explicitly: 
\begin{align}
&[ E, {Z}_\pm] =   {Z}_\pm, \quad [ {M}, {Z}_\pm] =  \pm {Z}_\pm, \quad [E,{W}] = 2 {W},\nonumber\\
&[Z_\pm,Z_{\pm}^{\dagger}] = -N,\qquad [Z_\pm,W^{\dagger}] = -Z_\mp^{\dagger},\nonumber\\
&[W,W^{\dagger}] = -E-N.\rule{0mm}{3.5mm}\label{ZWcomm}
\end{align}
A convenient combination is 
\beq
\tilde W=W-Z_{-}Z_{+}N^{-1}
\eeq
satisfying $[Z_\pm, \tilde W^{\dagger}]  = 0$. The operators $Z_+Z_+^\dagger$, $Z_-Z_-^\dagger$ and $\tilde W\tilde W^\dagger$ commute among themselves and with $N$, $E$, $M$ and $\mathcal{H}_{res}$. In fact, if one chooses a large $(N,E,M)$-block and diagonalizes $\mathcal{H}_{res}$, $Z_+Z_+^\dagger$, $Z_-Z_-^\dagger$ and $\tilde W\tilde W^\dagger$ simultaneously within this block, this allows for recovering the energy eigenvalues of all the lower blocks from which one can reach the given block by the moves in Fig.~\ref{GP2_raising}. Indeed, the kernel (\ref{kern3}), containing all the newly added energy eigenvalues not found in the lower blocks, evidently corresponds to zero eigenvalues of  $Z_+Z_+^\dagger$, $Z_-Z_-^\dagger$ and $\tilde W\tilde W^\dagger$, while all the other `descendant' energy eigenvalues that have arrived in the current block from the lower blocks are marked by positive eigenvalues of $Z_+Z_+^\dagger$, $Z_-Z_-^\dagger$ and $\tilde W\tilde W^\dagger$. These latter eigenvalues are easily obtained from the commutation relations (\ref{ZWcomm}), and encode the number of moves  in Fig.~\ref{GP2_raising} necessary to reach the current block from the block where a particular energy eigenvalue first appeared.

\begin{figure}[t!]
\includegraphics[scale=0.3]{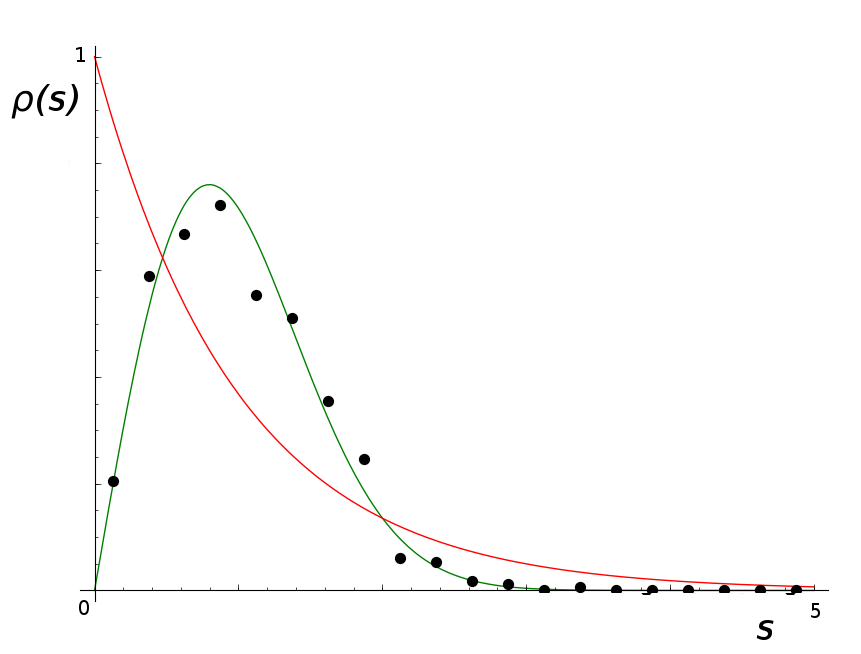}
\caption{The Wignerian distribution of the unfolded level spacings in the joint $(Z_+^\dagger,Z_-^\dagger,W^\dagger)$ kernel of the $(7,22,16)$-block of the resonant Hamiltonian. The green curve is the Wigner surmise (\ref{wign}), while the red curve is the Poisson distribution (\ref{pois}).}
\label{plot_GP2}
\end{figure}
In implementing our analysis of the level spacing statistics, as in section \ref{levelsp}, we must likewise separate the eigenvalues of $\mathcal{H}_{res}$ according to all the six mutually commuting conserved quantities $N$, $E$, $M$, $Z_+Z_+^\dagger$, $Z_-Z_-^\dagger$ and $\tilde W\tilde W^\dagger$. In practice, it is sufficient to consider zero values of the last three quantities, i.e., the eigenstates in the joint kernel (\ref{kern3}). Eigenvalues outside this kernel in a given $(N,E,M)$-block always descend through the action of $Z_\pm$ and $W$ from the kernel eigenvalues of a lower $(N,E,M)$-block.  With this picture in mind, we have plotted the unfolded level spacing distribution in the $(Z_\pm^\dagger,W^\dagger)$-kernel of the $(N,E,M)=(7,22,16)$ block, given in  Fig.~\ref{plot_GP2}. The distribution is Wignerian and suggests that no symmetry structures beyond (\ref{GP2_N}-\ref{GP2_W}) are to be expected. We give two extra level spacing distributions for different blocks in Appendix~\ref{App_B}. In this case, unlike the LLL sector, the distribution is very close to the Wigner-Dyson curve for all the blocks we have studied.


\section{Discussion}

We have analyzed the problem of energy level splitting for quantum bosons in a two-dimensional isotropic harmonic trap under the influence of weak contact interactions. The problem is shown to be identical (at linear order in the interaction strength) to diagonalizing the quantized resonant system of the Gross-Pitaevskii equation. Known symmetries of this resonant system impart a highly constrained structure to the energy spectrum. The action of the symmetries copies the energy shifts from lower to higher unperturbed energy levels, so that infinitely many exactly integer energy differences characteristic of noninteracting bosons in a harmonic trap survive (this is a reflection of the exact pattern in the spectrum valid at finite coupling). New values of energy shifts (in addition to the ones having descended from lower unperturbed levels via the action of the symmetries) always enter through explicitly described kernels of the symmetry generators. These symmetry structures can be used to optimize the numerical construction of eigenstate energies. The highest energy level splitting off the unperturbed level with $E$ units of energy and $N$ particles is explicitly given as
\beq
\tilde E_{max}= E+g\,\frac{N(N-1)}{2}+O(g^2),
\eeq
while all the others lie between $E$ and $\tilde E_{max}$.

We have furthermore analyzed the level spacing statistics in the fine structure emanating from individual unperturbed levels under the influence of weak contact interactions. The purpose is to use the standard lore of quantum chaos theory to shed light on the question of possible integrability (or, more generally, extra symmetry structures) in the resonant system of the Gross-Pitaevskii equation (whose quantized version describes the energy level splitting). Speculations about possible integrability of this resonant system (and its close relatives) have been voiced in the literature \cite{FGH, CF,quantres}, in particular due to some intriguing parallels seen between these systems and the cubic Szeg\H o equation \cite{GG}, which is known to be integrable. Our analysis of the level spacing statistics suggests that no integrability is to be expected. Outside the LLL sector (blocks with the maximal amount of rotation), we observe Wignerian level spacing distributions, characteristic of strongly chaotic systems. Within the LLL sector, there is a range of behaviors ranging from nearly Wignerian, to crossover, to Poissonian, which may be alluding to some very subtle analytic structure, but not to integrability.

A few classical systems possess symmetries closely analogous to those of the Gross-Pitaevskii equation, which is the classical version of (\ref{Horg}), both at the level of the original PDE, and at the level of extra symmetry enhancements in the resonant approximation. We specifically mention the conformally coupled cubic wave equation on the 3-sphere \cite{CF} and the Hartree equation in a harmonic trap in four spatial dimensions \cite{BEF}. Similarly, if one replaces the $|\Psi|^4$ interaction that has characterized our considerations by $|\Psi|^6$, there are systems in lower numbers of spatial dimensions \cite{fennell,quintic} that respect symmetry structures of the sort underlying our current study. There is additionally a very large class of resonant systems explicitly constructed in \cite{AO} in a way that guarantees the presence of the same type of symmetries. For all the systems mentioned, one expects that quantization will generate energy level patterns closely analogous to what we have observed in the present treatment.

We finally comment on the `quantum Gross-Pitaevskii equations' \cite{QGP}, a hierarchy of equations inspired by variational tensor network techniques and interpolating between the classical Gross-Pitaevskii equation and quantum bosons with contact interactions described by (\ref{Horg}). Since the operation of enhanced symmetries in the weak coupling regime has by now been seen both in the Gross-Pitaevskii equation \cite{continuous, GT, BBCE, BBCE2, GGT} and in the full quantum system in our present work, one expects it to have manifestations within the entire quantum Gross-Pitaevskii hierarchy. It would be interesting to investigate this subject further.


\begin{acknowledgments}

We thank Piotr Bizo\'n for a discussion on bounds for resonant Hamiltonians. This research has been supported by FWO-Vlaanderen (projects G044016N and G006918N), by Vrije Universiteit Brussel through the Strategic Research Program ``High-Energy Physics,'' and by CUniverse research promotion project (CUAASC) at Chulalongkorn University. M.~D.~C. is supported by a PhD fellowship from the Research Foundation Flanders (FWO).
\end{acknowledgments}

\onecolumngrid

\appendix

\section{Conservation of $Z_{+}$, $Z_{-}$ and $W$}\label{App_A}

The three proofs in this appendix closely retrace section 3 and the appendix of \cite{AO}. One has to prove that $[Z_\pm,\mathcal{H}_{res}]=[W,\mathcal{H}_{res}]=0$. Evaluating the commutators directly, which is identical to the time-differentiation in section 3 of \cite{AO}, produces expressions that vanish provided that certain four-term identities are satisfied by the interaction coefficients (\ref{intcoeff}). These identities are then proved using the following properties of the Laguerre polynomials:
\begin{align}
&\partial_\rho L_{n}^{\mu} = - L_{n-1}^{\mu+1},
\label{eq: derivative laguerre}\\
&L_{n}^{\mu} = L_{n}^{\mu+1}- L_{n-1}^{\mu+1},
\label{eq: difference laguerre}\\
&nL_{n}^{\mu} = (n+\mu)L_{n-1}^{\mu}- \rho L_{n-1}^{\mu+1}.
\label{eq: identity3 laguerre}
\end{align}

\subsection{Conservation of $Z_{+}$}

Direct computation of the commutator $[Z_+,\mathcal{H}_{res}]$ produces an expression that vanishes if
\begin{align}
\begin{split}
&\mathcal{D}_{nn_1n_2n_3}^{mm_1m_2m_3} \equiv \sqrt{\frac{n+m}{2}} \, C_{n-1,n_1n_2,n_3}^{m-1,m_1m_2,m_3} + \sqrt{\frac{n_1+m_1}{2}} \, C_{n,n_1-1,n_2n_3}^{m,m_1-1,m_2m_3} \\ &- \sqrt{\frac{n_2+m_2 + 2}{2}} \, C_{nn_1,n_2+1,n_3}^{mm_1,m_2+1,m_3} - \sqrt{\frac{n_3+m_3+2}{2}} \, C_{nn_1n_2,n_3+1}^{mm_1m_2,m_3+1} = 0,
\label{eq: four terms identity Z+}
\end{split}
\end{align}
whenever $n_3 = n+n_1 - n_2 -1$ and $m_3 = m+m_1 - m_2 -1$. Multiplying by
\beq
 \sqrt{\frac{((n+m)/2)!}{((n-m)/2)!}\frac{((n_1+m_1)/2)!}{((n_1-m_1)/2)!}\frac{((n_2+m_2)/2)!}{((n_2-m_2)/2)!}\frac{((n_3+m_3)/2)!}{((n_3-m_3)/2)!}}
\label{multproof}
\eeq
and substituting the interaction coefficients (\ref{CLaguerre}), we write
\begin{align}
\begin{split}
\mathcal{D}_{nn_1n_2n_3}^{\,mm_1m_2m_3} \,&\sim  \int^{\infty}_0 d\rho\, e^{-2\rho} \rho^{m+m_1-1} \Big[ \frac{n+m}{2} \, L_{\frac{n-m}{2}}^{m-1} \, L_{\frac{n_1-m_1}{2}}^{m_1}\, L_{\frac{n_2-m_2}{2}}^{m_2} \, L_{\frac{n_3-m_3}{2}}^{m_3} \\ &+ \frac{n_1+m_1}{2} \, L_{\frac{n-m}{2}}^{m} \, L_{\frac{n_1-m_1}{2}}^{m_1-1}\, L_{\frac{n_2-m_2}{2}}^{m_2} \, L_{\frac{n_3-m_3}{2}}^{m_3} - \rho L_{\frac{n-m}{2}}^{m} \, L_{\frac{n_1-m_1}{2}}^{m_1}\, L_{\frac{n_2-m_2}{2}}^{m_2+1} \, L_{\frac{n_3-m_3}{2}}^{m_3} \\ &- \rho L_{\frac{n-m}{2}}^{m} \, L_{\frac{n_1-m_1}{2}}^{m_1}\, L_{\frac{n_2-m_2}{2}}^{m_2} \, L_{\frac{n_3-m_3}{2}}^{m_3+1}\Big],
\label{eq: step1}
\end{split}
\end{align}
where we  have used $n_3 = n+n_1 - n_2 -1$ and $m_3 = m+m_1 - m_2 -1$.
We apply the identities (\ref{eq: difference laguerre}-\ref{eq: identity3 laguerre}) to the first two terms, while for the last two terms, we use (\ref{eq: difference laguerre}), obtaining
\begin{align}
\begin{split}
\mathcal{D}_{nn_1n_2n_3}^{\, mm_1m_2m_3}  &\sim \int^{\infty}_0 d\rho\, e^{-2\rho} \rho^{m+m_1-1} [ (m+m_1-2\rho) \, L_{\frac{n-m}{2}}^{m} \, L_{\frac{n_1-m_1}{2}}^{m_1}\, L_{\frac{n_2-m_2}{2}}^{m_2} \, L_{\frac{n_3-m_3}{2}}^{m_3} \\ &-\rho (  L_{\frac{n-m}{2}-1}^{m+1} \, L_{\frac{n_1-m_1}{2}}^{m_1}\, L_{\frac{n_2-m_2}{2}}^{m_2} \, L_{\frac{n_3-m_3}{2}}^{m_3} + L_{\frac{n-m}{2}}^{m} \, L_{\frac{n_1-m_1}{2}-1}^{m_1+1}\, L_{\frac{n_2-m_2}{2}}^{m_2} \, L_{\frac{n_3-m_3}{2}}^{m_3} \\ &+ L_{\frac{n-m}{2}}^{m} \, L_{\frac{n_1-m_1}{2}}^{m_1}\, L_{\frac{n_2-m_2}{2}-1}^{m_2+1} \, L_{\frac{n_3-m_3}{2}}^{m_3}+L_{\frac{n-m}{2}}^{m} \, L_{\frac{n_1-m_1}{2}}^{m_1}\, L_{\frac{n_2-m_2}{2}}^{m_2} \, L_{\frac{n_3-m_3}{2}-1}^{m_3+1})].
\end{split}
\end{align}
Using the remaining identity (\ref{eq: derivative laguerre}), we note that this expression is a total derivative:
\begin{align}
\begin{split}
\mathcal{D}_{nn_1n_2n_3}^{ \, mm_1m_2m_3}  &\sim \int^{\infty}_{0} d\rho\, \partial_{\rho} \left( e^{-2\rho}\rho^{m+m_1} L_{\frac{n-m}{2}}^{m} \, L_{\frac{n_1-m_1}{2}}^{m_1}\, L_{\frac{n_2-m_2}{2}}^{m_2} \, L_{\frac{n_3-m_3}{2}}^{m_3} \right) \\  
&=  e^{-2\rho}\rho^{m+m_1} L_{\frac{n-m}{2}}^{m} (\rho)\, L_{\frac{n_1-m_1}{2}}^{m_1} (\rho)\, L_{\frac{n_2-m_2}{2}}^{m_2}  (\rho)\, L_{\frac{n_3-m_3}{2}}^{m_3}  (\rho)\Big|_{\rho = 0}^{\rho = \infty}.
\end{split}
\end{align}
The Laguerre polynomials of the form $L_{\frac{n-m}{2}}^m$ do not have any powers of $\rho$ below $\rho^{-m}$, which ensures this last expression is zero, completing the proof.

 \subsection{Conservation of $Z_{-}$}
The  proof is analogous to the one above. Direct computation of the commutator $[Z_-,\mathcal{H}_{res}]$ produces an expression that vanishes if 
\begin{align}
\begin{split}
&\mathcal{D}_{nn_1n_2n_3}^{mm_1m_2m_3} \equiv 
\sqrt{\frac{n-m}{2}} \, C_{n-1,n_1n_2,n_3}^{m+1,m_1m_2,m_3} + \sqrt{\frac{n_1-m_1}{2}} \, C_{n,n_1-1,n_2n_3}^{m,m_1+1,m_2m_3} \\ &- \sqrt{\frac{n_2-m_2 + 2}{2}} \, C_{nn_1,n_2+1,n_3}^{mm_1,m_2-1,m_3} - \sqrt{\frac{n_3-m_3+2}{2}} \, C_{nn_1n_2,n_3+1}^{mm_1m_2,m_3-1} = 0,
\label{eq: four terms identity Z-}
\end{split}
\end{align}
whenever $n_3 = n+n_1 - n_2 -1$ and $m_3 = m+m_1 - m_2 +1$. Multiplying by (\ref{multproof}) and substituting (\ref{CLaguerre}), we get
\begin{align*}
\mathcal{D}_{nn_1n_2n_3}^{\, mm_1m_2m_3} \,  &\sim  \int^{\infty}_0 d\rho\, e^{-2\rho} \rho^{m+m_1} \Big[\rho \, L_{\frac{n-m}{2}-1}^{m+1} \, L_{\frac{n_1-m_1}{2}}^{m_1}\, L_{\frac{n_2-m_2}{2}}^{m_2} \, L_{\frac{n_3-m_3}{2}}^{m_3} +  \rho \, L_{\frac{n-m}{2}}^{m} \, L_{\frac{n_1-m_1}{2}-1}^{m_1+1}\, L_{\frac{n_2-m_2}{2}}^{m_2} \, L_{\frac{n_3-m_3}{2}}^{m_3} \\
&\hspace{-5mm}- \frac{n_2-m_2+2}{2} L_{\frac{n-m}{2}}^{m} \, L_{\frac{n_1-m_1}{2}}^{m_1}\, L_{\frac{n_2-m_2}{2}+1}^{m_2-1} \, L_{\frac{n_3-m_3}{2}}^{m_3} - \frac{n_3-m_3+2}{2} L_{\frac{n-m}{2}}^{m} \, L_{\frac{n_1-m_1}{2}}^{m_1}\, L_{\frac{n_2-m_2}{2}}^{m_2} \, L_{\frac{n_3-m_3}{2}+1}^{m_3-1}\Big].
\end{align*}
We apply identity (\ref{eq: derivative laguerre}) on the first two terms. The last two terms can be rewritten by applying (\ref{eq: difference laguerre}) first, followed by (\ref{eq: identity3 laguerre}) and using again  (\ref{eq: difference laguerre}), which is equivalent to
\beq
\left( \frac{n-m}{2} +1 \right) L_{\frac{n-m}{2} +1}^{m-1} = (m - \rho ) L_{\frac{n-m}{2}}^{m} - \rho L_{\frac{n-m}{2}-1}^{m+1}.
\eeq
Afterwards, we group all the terms to isolate a total derivative, as in the previous proof
\begin{align}
\begin{split}
\mathcal{D}_{nn_1n_2n_3}^{\, mm_1m_2m_3} \,  &\sim  -\int^{\infty}_0 d\rho\, e^{-2\rho} \rho^{m+m_1} [ (m_2+m_3 - 2\rho) \, L_{\frac{n-m}{2}}^{m} \, L_{\frac{n_1-m_1}{2}}^{m_1}\, L_{\frac{n_2-m_2}{2}}^{m_2} \, L_{\frac{n_3-m_3}{2}}^{m_3} \\ &+ \rho \partial_\rho(\, L_{\frac{n-m}{2}}^{m} \, L_{\frac{n_1-m_1}{2}}^{m_1}\, L_{\frac{n_2-m_2}{2}}^{m_2} \, L_{\frac{n_3-m_3}{2}}^{m_3} )] \\
&=  -\int^{\infty}_{0} d\rho\, \partial_{\rho} \left( e^{-2\rho}\rho^{m+m_1+1} L_{\frac{n-m}{2}}^{m} \, L_{\frac{n_1-m_1}{2}}^{m_1}\, L_{\frac{n_2-m_2}{2}}^{m_2} \, L_{\frac{n_3-m_3}{2}}^{m_3} \right) \\  
&=  e^{-2\rho}\rho^{m+m_1+1} L_{\frac{n-m}{2}}^{m}(\rho) \, L_{\frac{n_1-m_1}{2}}^{m_1}(\rho)\, L_{\frac{n_2-m_2}{2}}^{m_2}(\rho)\, L_{\frac{n_3-m_3}{2}}^{m_3}(\rho) \Big|_{\rho = 0}^{\rho = \infty}.
\end{split}
\end{align}
The expression is zero, which proves the conservation of $Z_{-}$.

\subsection{Conservation of $W$}

Direct computation of the commutator $[W,\mathcal{H}_{res}]$ produces an expression that vanishes if 
\begin{align}
\begin{split}
&\mathcal{D}_{nn_1n_2n_3}^{mm_1m_2m_3} \equiv \frac{\sqrt{(n+2)^2-m^2}}{2} \, C_{n+2,n_1n_2,n_3}^{m,m_1m_2,m_3} + \frac{\sqrt{(n_1+2)^2-m_1^2}}{2} \, C_{n,n_1+2,n_2n_3}^{m,m_1,m_2m_3} \\ &-  \frac{\sqrt{n_2^2-m_2^2}}{2}\, C_{nn_1,n_2-2,n_3}^{mm_1,m_2,m_3} - \frac{\sqrt{n_3^2-m_3^2}}{2} \, C_{nn_1n_2,n_3-2}^{mm_1m_2,m_3} = 0,
\label{eq: four terms identity W}
\end{split}
\end{align}
whenever $ n+n_1 =  n_2  +n_3 - 2$ and $m + m_1 = m_2+m_3$. Multiplying by (\ref{multproof}) and substituting (\ref{CLaguerre}), we get
\begin{align}
\begin{split}
&\mathcal{D}_{nn_1n_2n_3}^{\,mm_1m_2m_3} \, \sim  \int^{\infty}_0 d\rho\, e^{-2\rho} \rho^{m+m_1} \Big[ \frac{n+2-m}{2} \, L_{\frac{n+2-m}{2}}^{m} \, L_{\frac{n_1-m_1}{2}}^{m_1}\, L_{\frac{n_2-m_2}{2}}^{m_2} \, L_{\frac{n_3-m_3}{2}}^{m_3} \\ &+ \frac{n_1+2-m_1}{2} \, L_{\frac{n-m}{2}}^{m} \, L_{\frac{n_1+2-m_1}{2}}^{m_1}\, L_{\frac{n_2-m_2}{2}}^{m_2} \, L_{\frac{n_3-m_3}{2}}^{m_3} - \frac{n_2+m_2}{2} L_{\frac{n-m}{2}}^{m} \, L_{\frac{n_1-m_1}{2}}^{m_1}\, L_{\frac{n_2-2-m_2}{2}}^{m_2} \, L_{\frac{n_3-m_3}{2}}^{m_3} \\ &- \frac{n_3+m_3}{2} L_{\frac{n-m}{2}}^{m} \, L_{\frac{n_1-m_1}{2}}^{m_1}\, L_{\frac{n_2-m_2}{2}}^{m_2} \, L_{\frac{n_3-2-m_3}{2}}^{m_3}\Big]
\label{eq: step1}
\end{split}
\end{align}
We now apply the identity (\ref{eq: identity3 laguerre}) to the four terms, yielding
\begin{align}
\begin{split}
\mathcal{D}_{nn_1n_2n_3}^{\, mm_1m_2m_3}  &\sim \int^{\infty}_0 d\rho\, e^{-2\rho} \rho^{m+m_1} [ (m+m_1+1) \, L_{\frac{n-m}{2}}^{m} \, L_{\frac{n_1-m_1}{2}}^{m_1}\, L_{\frac{n_2-m_2}{2}}^{m_2} \, L_{\frac{n_3-m_3}{2}}^{m_3} \\ &-\rho (  L_{\frac{n-m}{2}}^{m+1} \, L_{\frac{n_1-m_1}{2}}^{m_1}\, L_{\frac{n_2-m_2}{2}}^{m_2} \, L_{\frac{n_3-m_3}{2}}^{m_3} + L_{\frac{n-m}{2}}^{m} \, L_{\frac{n_1-m_1}{2}}^{m_1+1}\, L_{\frac{n_2-m_2}{2}}^{m_2} \, L_{\frac{n_3-m_3}{2}}^{m_3} \\ &+ L_{\frac{n-m}{2}}^{m} \, L_{\frac{n_1-m_1}{2}}^{m_1}\, L_{\frac{n_2-m_2}{2}-1}^{m_2+1} \, L_{\frac{n_3-m_3}{2}}^{m_3}+L_{\frac{n-m}{2}}^{m} \, L_{\frac{n_1-m_1}{2}}^{m_1}\, L_{\frac{n_2-m_2}{2}}^{m_2} \, L_{\frac{n_3-m_3}{2}-1}^{m_3+1})],
\end{split}
\end{align}
where we have  used $ n+n_1 =  n_2  +n_3 - 2$ and $m + m_1 = m_2+m_3$.
Finally, we apply (\ref{eq: difference laguerre}) to the second and third term, and identify a total derivative using (\ref{eq: derivative laguerre}), giving
\begin{align}
\begin{split}
\mathcal{D}_{nn_1n_2n_3}^{ \, mm_1m_2m_3}  &\sim \int^{\infty}_{0} d\rho\, \partial_{\rho} \left( e^{-2\rho}\rho^{m+m_1+1} L_{\frac{n-m}{2}}^{m} \, L_{\frac{n_1-m_1}{2}}^{m_1}\, L_{\frac{n_2-m_2}{2}}^{m_2} \, L_{\frac{n_3-m_3}{2}}^{m_3} \right) \\  
&=  e^{-2\rho}\rho^{m+m_1+1} L_{\frac{n-m}{2}}^{m}(\rho) \, L_{\frac{n_1-m_1}{2}}^{m_1}(\rho)\, L_{\frac{n_2-m_2}{2}}^{m_2}(\rho) \, L_{\frac{n_3-m_3}{2}}^{m_3}(\rho) \Big|_{\rho = 0}^{\rho = \infty}.
\end{split}
\end{align}
The Laguerre polynomials of the form $L_{\frac{n-m}{2}}^m$ do not have any powers of $\rho$ below $\rho^{-m}$, which ensures that the last expression is zero and $W$ is conserved.


\section{More information on the level spacing statistics} \label{App_B}

In this appendix, we present a few extra plots of level spacing statistics that are not crucial for our conclusions in the main text (that integrability is unlikely), but give a broader perspective, and also bring forth some interesting details.

	\begin{figure}[t]
		\subfloat[$N=55,\,\,E=30$.]{\includegraphics[scale=0.35]{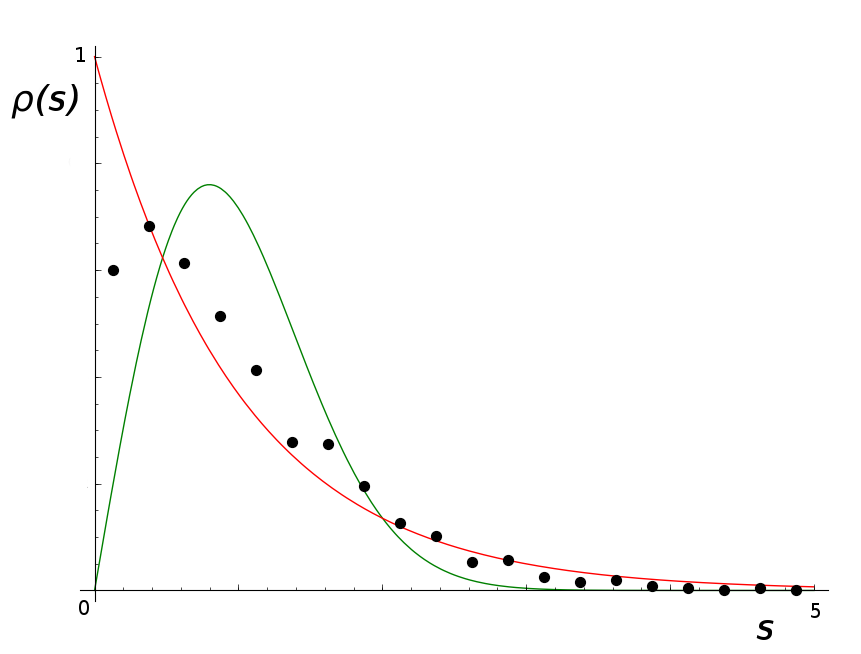}}
		\subfloat[$N=7,\,\,E=43$.]{\includegraphics[scale=0.35]{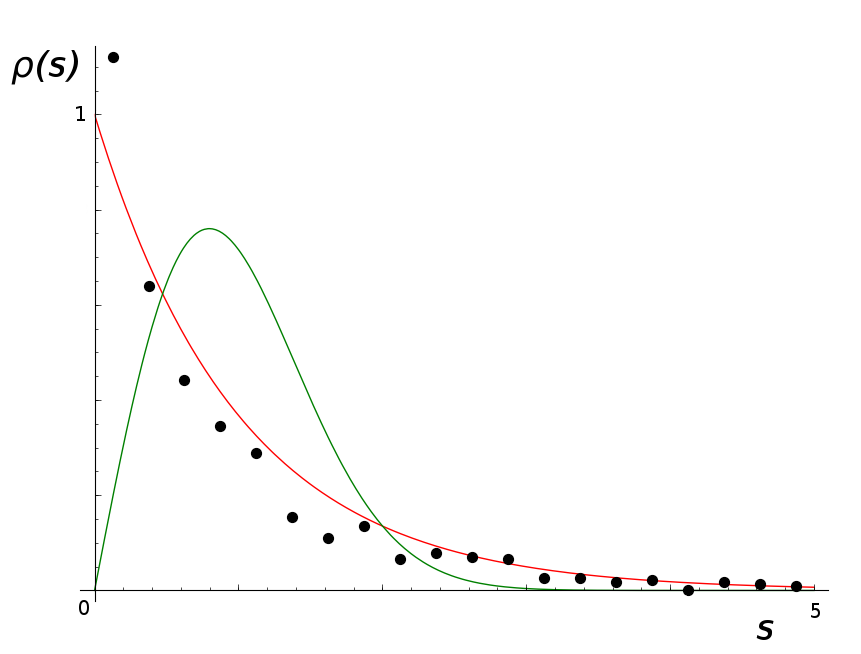}}
		\caption{Crossover (a) and Poissonian (b) level spacing distributions for two different blocks in the LLL sector.}
		\label{LLLextra}
	\end{figure}
	\begin{figure}[t]
		\subfloat[$N=10,\,\,E=16,\,\,M=2$.]{\includegraphics[scale=0.35]{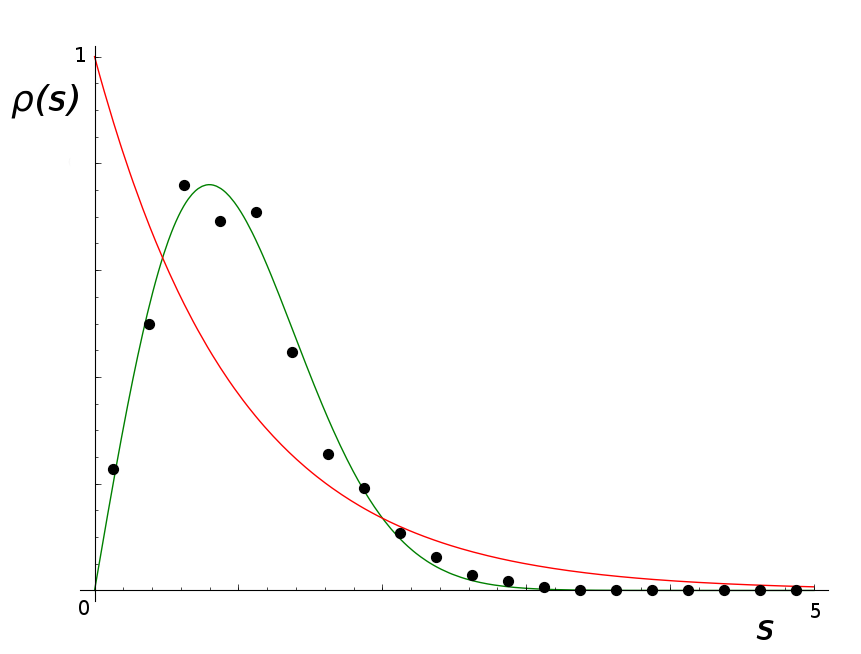}}
		\subfloat[$N=10,\,\,E=17,\,\,M=3$.]{\includegraphics[scale=0.35]{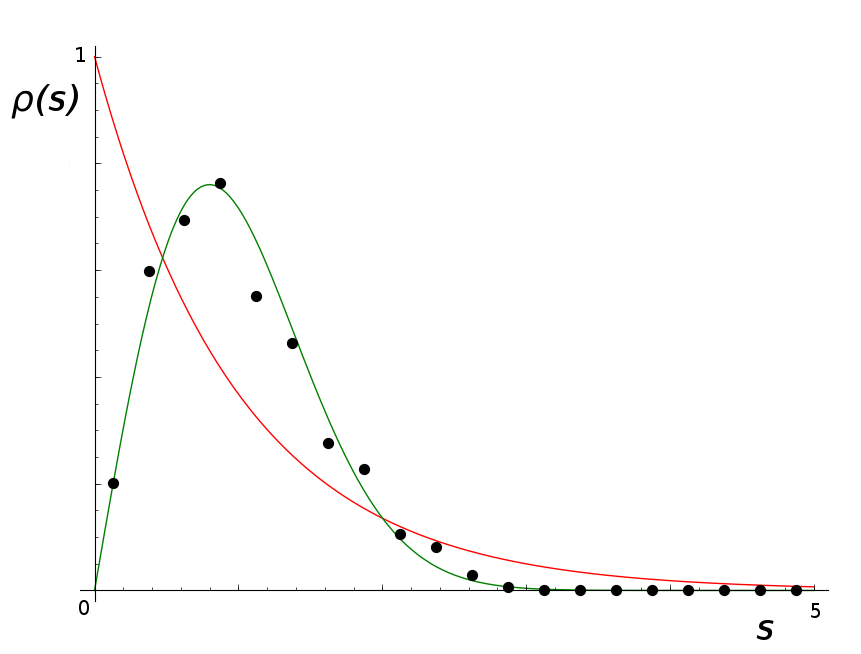}}
		\caption{Wignerian distributions in two different blocks away from the LLL sector.}
		\label{GP2extra}
	\end{figure}
As we remarked in the main text in section \ref{section_LLL}, the LLL sector is characterized by distributions that are not exactly Wignerian, but somewhat shifted towards the Poissonian curve. Such distributions are often described as `crossover' behaviors, and characterize chaotic systems with relatively weak chaotic dynamics (for example, small chaotic perturbations of integrable systems). Observing these distributions makes integrability unlikely, but leaves some room for more subtle features. The distributions appear close to Wignerian when $N\approx M$ and morph in the direction of the Poissonian curve as one moves away from that region. If $N\ll M$, Poissonian behaviors are observed. This is depicted in Fig.~\ref{LLLextra} for two particular blocks.

Away from the LLL sector, we have consistently observed Wignerian distributions, exemplified by two specific blocks in Fig.~\ref{GP2extra}. In particular, the level repulsion is clearly visible as the leftmost point 
of each plot fits the Wignerian curve very accurately.


\section{Bounds on the two-dimensional resonant Hamiltonian} \label{App_C}

Proofs of bounds on the resonant Hamiltonian (\ref{resH}) are given in \cite{FGH}, but in a language that may be not immediately accessible for readers with physics backgrounds. The purpose of this appendix is to recast these proofs in a way that is more straightforward, even if less rigorous. We also need to translate the proofs of \cite{FGH}, given for the classical theory, into statements about the quantum resonant Hamiltonian (\ref{resH}).

\subsection{Positivity}

Consider
\beq
\Psi(\bfr,\te)=\sum_{n,m}\al_{nm}\,\psi_{nm}(\bfr)\,e^{-i(n+1)\te},
\label{psiHeisenbrg}
\eeq
with the normalized harmonic oscillator energy eigenstates $\psi_{nm}$ given by (\ref{psipossgn}) and $\bfr\equiv(x,y)$. This expression is, of course, simply the Heisenberg operator of the field $\Psi$. We use the angle-like notation $\te$ for the time variable in this representation to emphasize that the harmonic oscillator evolution (and hence the above Heisenberg operator) is exactly periodic. With this notation, the resonant Hamiltonian (\ref{resH}) with the interaction coefficients (\ref{intcoeff}) is written as
\beq
\mathcal{H}_{res}=\frac12 \int_{-\pi}^{\pi} \Psi^{\dagger 2}(\bfr,\te)\,\Psi^2(\bfr,\te) \,\dr\,d\te,
\label{resHeisenbrg}
\eeq
where we have used $\int_{-\pi}^{\pi} e^{i(n_1+n_2-n_3-n_4)\te} d\te=2\pi\de_{n_1+n_2,n_3+n_4}$.
Since this expression is an integral of the manifestly positive operator $\Psi^{\dagger 2}\Psi^2$, $\mathcal{H}_{res}$ is positive.

\subsection{Upper bound}

Proving the upper bound $\mathcal{H}_{res}\le N(N-1)/2$ will require more effort than the positivity proof. The classical version of this statement is proved in \cite{FGH} and turns out equivalent to constructing bounds on the so-called Strichartz norms. The latter problem has been extensively studied in the mathematical literature starting with \cite{strnorm1,strnorm2}, however, sharp bounds with explicit numerical coefficients of the sort we need here appeared only rather recently \cite{strnorm3,strnorm4}. We find the proofs of \cite{strnorm4} particularly graceful and easy to adapt for our purposes. Arguments based on the heat kernel, as in \cite{FGH,strnorm3}, are not straightforwardly translated to the quantum Hamiltonian (\ref{resH}). Our presentation below essentially retraces the proof of \cite{strnorm4} in a physicist's language, and extends it to the quantum case.

The evolution of (\ref{psiHeisenbrg}) is expressed through the harmonic oscillator propagator as
\beq
\Psi(\bfr,\te)=e^{i\te\left(\del_x^2+\del_y^2-x^2-y^2\right)/2}\Psi(\bfr,0) =\frac{1}{2\pi i \sin \te}\int \exp\left[-\frac{(\bfr^2+\bfr'^{2})\cos\te-2(\bfr\bfr')}{2i\sin\te}\right]\Psi(\bfr',0) \,\dr'.
\label{Psievol}
\eeq
We note that $\Psi(\bfr,\te+\pi)=-\Psi(-\bfr,\te)$, which lets us reduce the range of integration in (\ref{resHeisenbrg}) as
\beq
\mathcal{H}_{res}=\int_{0}^{\pi} \Psi^{\dagger 2}(\bfr,\te)\,\Psi^2(\bfr,\te) \,\dr\,d\te.
\label{resHsbhalf}
\eeq
Then,
\beq
\mathcal{H}_{res}=\int_{0}^{\pi}d\te\int \dr\left(\prod_{i=1}^4\dr_i \right)\frac{e^{\frac{i}2(\bfr_3^2+\bfr_4^2-\bfr_1^2-\bfr_2^2)\cot\te}e^{i(\bfr(\bfr_1+\bfr_2-\bfr_3-\bfr_4))/\sin\te}}{(2\pi\sin\te)^4}\Psi^{\dagger}(\bfr_1)\Psi^{\dagger}(\bfr_2)\Psi(\bfr_3)\Psi(\bfr_4),
\eeq
where we have reverted to the notation $\Psi(\bfr)\equiv \Psi(\bfr,0)$.
Now introduce ${\bf k}=\bfr/\sin\te$ and $\tau=\cot\te$ so that
\begin{align}
\mathcal{H}_{res}&=\int_{-\infty}^{\infty}d\tau\int d{\bf k}\left(\prod_{i=1}^4\dr_i \right)\frac{e^{i\tau(\bfr_3^2+\bfr_4^2-\bfr_1^2-\bfr_2^2)/2}e^{i({\bf k}(\bfr_1+\bfr_2-\bfr_3-\bfr_4))}}{(2\pi)^4}\Psi^{\dagger}(\bfr_1)\Psi^{\dagger}(\bfr_2)\Psi(\bfr_3)\Psi(\bfr_4)\nonumber\\
&=\frac{1}{\pi}\int\left( \prod_{i=1}^4\dr_i\right)\de(\bfr_3^2+\bfr_4^2-\bfr_1^2-\bfr_2^2)\,\de(\bfr_1+\bfr_2-\bfr_3-\bfr_4)\,\Psi^{\dagger}(\bfr_1)\Psi^{\dagger}(\bfr_2)\Psi(\bfr_3)\Psi(\bfr_4).
\label{rescartdel}
\end{align}
 We note that the transformations we have just performed are closely linked to the connection between harmonic oscillator and free motion known as the lens transform \cite{Carles,Tao}. One may notice the appearance of the characteristic energy-momentum conservation of a free particle in the $\de$-functions of (\ref{rescartdel}). It may seem surprising that the `energies' and `momenta' in the $\de$-functions are expressed through the coordinates $\bfr_i$, but in fact the expression is invariant under Fourier transform, as emphasized in \cite{FGH}, so the momentum-space analog of (\ref{rescartdel}) is given by the same expression, but with $\bfr_i$ replaced by the corresponding momenta ${\bf k}_i$ and the conservation of energy given by the standard expression. Note that the said invariance of $\mathcal{H}_{res}$ under the Fourier transform of $\Psi$ is a direct consequence of (\ref{resHsbhalf}) being an integral of a periodic function, whose integration range can be freely shifted, and (\ref{Psievol}) being invariant under a shift of $\te$ by $\pi/2$ combined with a Fourier transform (which is the well-known emergence of Fourier transforms at a quarter-period in the evolution of the quantum harmonic oscillator).

Back to our problem, introduce $\bfR=(\bfr_1+\bfr_2)/2$, $\bfs=(\bfr_1-\bfr_2)/2$, $\bfR'=(\bfr_3+\bfr_4)/2$, $\bfs'=(\bfr_3-\bfr_4)/2$ so that $\dr_1\dr_2\dr_3\dr_4=4\dR\,\ds\,\dR'\,\ds'$ and
\beq
\mathcal{H}_{res}=\frac{1}{\pi}\int \dR\,\ds\,\dR'\,\ds'\, \,\de(\bfs^2-\bfs'^2)\,\de(\bfR-\bfR')\,\Psi^{\dagger}(\bfR+\bfs)\,\Psi^{\dagger}(\bfR-\bfs)\,\Psi(\bfR'+\bfs')\,\Psi(\bfR'-\bfs').
\label{HresRs}
\eeq
We now examine the action of the following operator $\hat S$ on a general (complex number valued) test function $h(\bfR,\bfs)$:
\beq
[\hat S h](\bfR,\bfs)=\frac{1}{\pi}\int \dR'\,\ds'\, \,\de(\bfs^2-\bfs'^2)\,\de(\bfR-\bfR')\,h(\bfR',\bfs').
\eeq
In polar coordinates $(s,\s)$ on the $\bfs$-plane, any function can be expanded in a Fourier series as
\beq
h(\bfR,\bfs)=\sum_{m=-\infty}^\infty h_m(\bfR,s) \, e^{im\s}.
\eeq
Then,
\beq
[\hat S h](\bfR,\bfs)=\frac{1}{\pi}\sum_{m=-\infty}^\infty h_m(\bfR,s) \int_0^{2\pi}e^{im\s}d\s \int_0^\infty \de(s^2-s'^2)\,s'\,ds'= h_0(\bfR,s).
\eeq
Thus, $\hat S$ is simply the orthogonal projector on the zero angular momentum subspace ($m=0$). But any projector is bounded from above by the identity operator, and hence, for any $h(\bfR,\bfs)$,
\beq
\frac{1}{\pi}\int \dR\,\ds\,\ds'\, \,\de(\bfs^2-\bfs'^2)\, h^*(\bfR,\bfs) h(\bfR,\bfs')\le \int \dR\,\ds\, |h(\bfR,\bfs)|^2.
\label{hbound}
\eeq

Consider the expectation value $\langle \Phi|\mathcal{H}_{res}|\Phi\rangle$ with an arbitrary state $|\Phi\rangle$.
Insert the decomposition of unity in terms of a complete basis of states $|\Phi'\rangle$ into $\mathcal{H}_{res}$ given by (\ref{HresRs}) between $\Psi^{\dagger}(\bfR+\bfs)\,\Psi^{\dagger}(\bfR-\bfs)$ and 
$\Psi(\bfR'+\bfs')\,\Psi(\bfR'-\bfs')$ to obtain
\begin{align}
\langle \Phi|\mathcal{H}_{res}|\Phi\rangle&= \sum_{|\Phi'\rangle} \frac{1}{\pi}\int \dR\,\ds\,\ds'\, \,\de(\bfs^2-\bfs'^2)\langle \Phi|\Psi^{\dagger}(\bfR+\bfs)\Psi^{\dagger}(\bfR-\bfs)|\Phi'\rangle\langle \Phi'|\Psi(\bfR+\bfs')\Psi(\bfR-\bfs')|\Phi\rangle\nonumber\\
&= \sum_{|\Phi'\rangle}  \frac{1}{\pi}\int \dR\,\ds\,\ds'\, \,\de(\bfs^2-\bfs'^2)\,h_{\Phi\Phi'}^*(\bfR,\bfs) h_{\Phi\Phi'}(\bfR,\bfs')\label{psipsi}
\end{align}
with $h_{\Phi\Phi'}(\bfR,\bfs)=\langle \Phi'|\Psi(\bfR+\bfs)\Psi(\bfR-\bfs)|\Phi\rangle$. By (\ref{hbound}), for any $|\Phi\rangle$,
\begin{align}
\langle \Phi|\mathcal{H}_{res}|\Phi\rangle\le \sum_{|\Phi'\rangle}\int \dR\,\ds\, \big|h_{\Phi\Phi'}(\bfR,\bfs)\big|^2 &=  \sum_{|\Phi'\rangle}\int \dR\,\ds\langle \Phi|\Psi^{\dagger}(\bfR+\bfs)\Psi^{\dagger}(\bfR-\bfs)|\Phi'\rangle\langle \Phi'|\Psi(\bfR+\bfs)\Psi(\bfR-\bfs)|\Phi\rangle\nonumber\\
&=\langle \Phi|\int \dR\,\ds\,\Psi^{\dagger}(\bfR+\bfs)\Psi^{\dagger}(\bfR-\bfs)\Psi(\bfR+\bfs)\Psi(\bfR-\bfs)|\Phi\rangle.
\end{align}
Converting back from $\bfR$ and $\bfs$ to $\bfr_1$ and $\bfr_2$, we get
\beq
\int \dR\,\ds\,\Psi^{\dagger}(\bfR+\bfs)\Psi^{\dagger}(\bfR-\bfs)\Psi(\bfR+\bfs)\Psi(\bfR-\bfs)=\frac12\int \dr_1\dr_2\, \Psi^{\dagger}(\bfr_1)\Psi^{\dagger}(\bfr_2)\Psi(\bfr_1)\Psi(\bfr_2)=\frac{N(N-1)}2,
\eeq
since $N=\int\Psi^\dagger(\bfr)\Psi(\bfr)\dr$ and $[\Psi^\dagger(\bfr'),\Psi(\bfr)]=-\de(\bfr'-\bfr)$.
We have thus proved that $\mathcal{H}_{res}$ is bounded from above by $N(N-1)/2$, as intended.

We conclude with some remarks on the proof of the upper bound we have just presented. Our proof is admittedly rather naive from a mathematical perspective. First, we have not been careful in specifying the functional spaces to which the various functions belong in deriving (\ref{hbound}), though a mathematically rigorous derivation of the same bound can be found in \cite{strnorm4}. Second, in the quantum context, everything becomes even more subtle, as rigorous analysis would require a much more accurate definition of the space of states and dealing with issues of convergence in the decomposition of unity in (\ref{psipsi}), etc. We believe that our derivations capture the essence of the problem, however. Importantly, the end result is just a constraint on the diagonalization of finite-sized matrices for the problem of level splitting found in the main text, which is perfectly well-defined in terms of elementary mathematics. This constraint (the eigenvalues lying between $0$ and $N(N-1)/2$) is furthermore empirically validated by our numerical diagonalization for concrete energy levels.

The derivations presented above are phrased through the representation of the resonant Hamiltonian given by (\ref{rescartdel}), which is illuminating, but rather unintuitive in the context of a harmonic trap problem. It would be interesting to construct a more direct proof in the Laguerre polynomial basis, but we are not aware of an easy way to do this. If one only keeps the modes with $m=0$ in the resonant system, the positivity results for integrals of products of Laguerre polynomials \cite{lagpos} and the summation identities of \cite{BBCE2} allow for a direct generalization of the elementary upper bound proof in section \ref{boundLLL}. This proof cannot be immediately adapted, however, to the full resonant system, which has both positive and negative interaction coefficients.

\twocolumngrid


\begin{thebibliography}{99}

\bibitem{split1}T.~Busch, B.-G. Englert, K.~Rz\k{a}\.zewski and M.~Wilkens, \emph{Two cold atoms in a harmonic trap,} Found. Phys. {\bf 28} (1998) 549.

\bibitem{split2}M.~Block and M.~Holthaus, \emph{Pseudopotential approximation in a harmonic trap,} Phys.\ Rev.\ A {\bf 65} (2002) 052102.

\bibitem{split3}F.~Werner and Y.~Castin,  \emph{The unitary three-body problem in a trap,} Phys.\ Rev.\ Lett.\ {\bf 97} (2006) 150401 \arXiv{cond-mat/0507399}; \emph{The unitary gas in an isotropic harmonic trap: symmetry properties and applications,} Phys.\ Rev.\ A {\bf 74} (2006) 053604 \arXiv{cond-mat/0607821}.

\bibitem{split4}P.~Shea, B.~P.~van Zyl and R.~K.~Bhaduri,  \emph{The two-body problem of ultra-cold atoms in a harmonic trap,} Am. J. Phys. {\bf 77}  (2009) 511  	\arXiv{0807.2979} [physics.atom-ph].

\bibitem{split5}N.~L.~Harshman, \emph{Symmetries of three harmonically trapped particles in one dimension}, Phys. Rev. A {\bf 86} (2012) 052122 \arXiv{1209.1398} [quant-ph].

\bibitem{split6}M.~A.~Garc\'\i a-March, B.~Juli\'a-D\'\i az, G.~E.~Astrakharchik, J.~Boronat and A. Polls, \emph{Distinguishability, degeneracy and correlations in three harmonically trapped bosons in one-dimension,} Phys.\ Rev.\ A {\bf 90} (2014) 063605 \arXiv{1410.7307} [cond-mat.quant-gas].

\bibitem{split7}P.~Mujal, E.~Sarl\'e, A.~Polls and B.~Juli\'a-D\'\i az, \emph{Quantum correlations and degeneracy of identical bosons in a two-dimensional harmonic trap,} Phys.\ Rev.\ A {\bf 96} (2017) 043614 \arXiv{1707.04166} [cond-mat.quant-gas].

\bibitem{split8}P.~Ko\'scik and T.~Sowi\'nski, \emph{Exactly solvable model of two trapped quantum particles interacting via finite-range soft-core interactions,} Sci.\ Rep. {\bf 8} (2018) 48 \arXiv{1707.04240} [cond-mat.quant-gas].

\bibitem{split9}D.~Blume, M.~W.~C.~Sze and  J.~ L.~Bohn, \emph{Harmonically trapped four-boson system,} Phys.\ Rev.\ A {\bf 97} (2018) 033621 \arXiv{1802.00129} [cond-mat.quant-gas].

\bibitem{breathing}L.~P.~Pitaevskii and A.~Rosch, \emph{Breathing modes and hidden symmetry of trapped atoms in 2d,} Phys.\ Rev.\ A {\bf 55} (1997) R835 \arXiv{cond-mat/9608135}.

\bibitem{FGH}
E.~Faou, P.~Germain and Z.~Hani, {\em The weakly nonlinear large box limit of the 2d cubic nonlinear Schr\"odinger equation,} J. Amer. Math. Soc. {\bf 29} (2016) 915 \arXiv{1308.6267} [math.AP].

\bibitem{continuous}
P.~Germain, Z.~Hani and L.~Thomann, {\em On the continuous resonant equation for NLS: I. Deterministic analysis.}
J.\ Math.\ Pur.\ Appl. {\bf 105} (2016) 131,  \arXiv{1501.03760} [math.AP].

\bibitem{quantres}
  O.~Evnin and W.~Piensuk,
  {\em Quantum resonant systems, integrable and chaotic,} J.\ Phys.\ A {\bf 52} (2019) 025102
 \arXiv{1808.09173} [math-ph].

\bibitem{GMW}
  T.~Guhr, A.~M\"uller-Groeling and H.~A.~Weidenm\"uller,
  \emph{Random matrix theories in quantum physics: common concepts},
  Phys.\ Rept.\  {\bf 299} (1998) 189
  \arXiv{cond-mat/9707301}.

\bibitem{haake} F.~Haake, \emph{Quantum signatures of chaos}, Springer (2001).

\bibitem{DKPR}
  L.~D'Alessio, Y.~Kafri, A.~Polkovnikov and M.~Rigol,
  \emph{From quantum chaos and eigenstate thermalization to statistical mechanics and thermodynamics,}
  Adv.\ Phys.\  {\bf 65} (2016) 239
  \arXiv{1509.06411} [cond-mat.stat-mech].

\bibitem{btint}M.~V.~Berry and M.~Tabor, \emph{Level clustering in the regular spectrum},
Proc.\ Roy.\ Soc.\ A {\bf 356} (1977) 375.

\bibitem{BGS} O.~Bohigas, M.-J.~Giannoni and C.~Schmit, \emph{Characterization of chaotic quantum
spectra and universality of level fluctuation laws,} Phys.\ Rev.\ Lett.\ {\bf 52} (1984) 1.

\bibitem{dahl}J.~P.~Dahl and W.~P.~Schleich, \emph{State operator, constants of the motion, and Wigner functions: The two-dimensional
isotropic harmonic oscillator},  Phys.\ Rev.\ A {\bf 79} (2009) 024101.

\bibitem{BBCE2}
  A.~Biasi, P.~Bizo\'n, B.~Craps and O.~Evnin,
 {\em Two infinite families of resonant solutions for the Gross-Pitaevskii equation,}
  Phys.\ Rev.\ E {\bf 98} (2018) 032222
  \arXiv{1805.01775} [cond-mat.quant-gas].

\bibitem{madagascar}
  O.~Evnin,
  {\em Spectroscopy instead of scattering: particle experimentation in AdS spacetime,} in Proceedings of HEPMAD18 (Antananarivo, Madagascar), eConf {\bf C180906} (2018)
  \arXiv{1812.07132} [hep-th].

\bibitem{BEL}
  B.~Craps, O.~Evnin and V.~Luyten,
  {\em Maximally rotating waves in AdS and on spheres,} JHEP {\bf 1709} (2017) 059
   \arXiv{1707.08501} [hep-th].

\bibitem{BEF}
  P.~Bizon, O.~Evnin and F.~Ficek,
  \emph{A nonrelativistic limit for AdS perturbations,}
  JHEP {\bf 1812} (2018) 113
  \arXiv{1810.10574} [gr-qc].
 
\bibitem{BSS}
  I.~Bertan, I.~Sachs and E.~D.~Skvortsov,
  {\em Quantum $\phi^4$ theory in AdS$_4$ and its CFT dual,}
 \arXiv{1810.00907} [hep-th].


\bibitem{CF}  P.~Bizo\'n, B.~Craps, O.~Evnin, D.~Hunik, V.~Luyten and M.~Maliborski,
 \emph{Conformal flow on $S^3$ and weak field integrability in AdS$_4$,} Comm.\ Math.\ Phys.\  {\bf 353} (2017) 1179 \arXiv{1608.07227} [math.AP].

\bibitem{BHP} P.~Bizo\'n, D.~Hunik-Kostyra and D.~Pelinovsky,
 {\em Ground state of the conformal flow on S$^{\,3}$}, 
\arXiv{1706.07726} [math.AP];
  {\em Stationary states of the cubic conformal flow on S$^{\,3}$,}
  \arXiv{1807.00426} [math-ph].

\bibitem{FPU}
  V.~Balasubramanian, A.~Buchel, S.~R.~Green, L.~Lehner and S.~L.~Liebling,
 {\em Holographic thermalization, stability of anti-de Sitter space, and the Fermi-Pasta-Ulam paradox,}
  Phys.\ Rev.\ Lett.\  {\bf 113} (2014) 071601
  \arXiv{1403.6471} [hep-th].

\bibitem{CEV} B.~Craps, O.~Evnin and J.~Vanhoof,
\emph{Renormalization group, secular term resummation and AdS (in)stability,}
 JHEP {\bf 1410} (2014) 48
 \arXiv{1407.6273} [gr-qc];
\emph{Renormalization, averaging, conservation laws and AdS (in)stability,}
JHEP {\bf 1501} (2015) 108
\arXiv{1412.3249} [gr-qc].

\bibitem{BMR} P.~Bizo\'n, M.~Maliborski, A.~Rostworowski, 
\emph{Resonant dynamics and the instability of anti-de Sitter spacetime,} 
Phys.\ Rev.\ Lett. {\bf 115} (2015) 081103
    \arXiv{1506.03519} [gr-qc].

\bibitem{bbbb}I.~Bia\l ynicki-Birula and Z.~Bia\l ynicka-Birula, \emph{Center-of-mass motion in the many-body theory of Bose-Einstein condensates,} Phys.\ Rev.\ A {\bf 65} (2002) 063606.

\bibitem{Niederer}
U.~Niederer, {\em The maximal kinematical invariance group of the harmonic oscillator,} Helv. Phys. Acta 46 (1973) 191.

\bibitem{OFN}
  K.~Ohashi, T.~Fujimori and M.~Nitta,
  {\em Conformal symmetry of trapped Bose-Einstein condensates and massive Nambu-Goldstone modes,}
  Phys.\ Rev.\ A {\bf 96} (2017) 051601
  \arXiv{1705.09118} [cond-mat.quant-gas].

\bibitem{Carles} R.~Carles, \emph{Critical nonlinear Schr\"odinger equations with and without harmonic potential,} Math.\ Mod.\ Meth.\ Appl.\ Sci. {\bf 12} (2002) 1513 \arXiv{cond-mat/0112414}.

\bibitem{Tao}
T. Tao, \emph{A pseudoconformal compactification of the nonlinear Schr\"odinger equation and applications,} New York J. Math. \textbf{15} (2009) 265 \arXiv{math/0606254} [math.AP].

\bibitem{motzkin1}R.~Movassagh and P.~W.~Shor, \emph{Power law violation of the area law in quantum spin chains,} Proc.\ Natl.\ Acad.\ Sci.\ {\bf 113} (2016) 13278 \arXiv{1408.1657} [quant-ph].

\bibitem{motzkin2}F.~Sugino and P.~Padmanabhan,
  \emph{Area law violations and quantum phase transitions in modified Motzkin walk spin chains,}
  J.\ Stat.\ Mech.\  {\bf 1801} (2018) 013101
  \arXiv{1710.10426} [quant-ph].

\bibitem{GT} P.~Germain and L.~Thomann,  \emph{On the high frequency limit of the LLL equation}, Quart. Appl. Math. {\bf 74} (2016) 633 \arXiv{1509.09080} [math.AP].

\bibitem{BBCE}
  A.~Biasi, P.~Bizo\'n, B.~Craps and O.~Evnin,
  {\em Exact lowest-Landau-level solutions for vortex precession in Bose-Einstein condensates,}
  Phys.\ Rev.\ A {\bf 96} (2017) 053615
  \arXiv{1705.00867} [cond-mat.quant-gas].

\bibitem{GGT}
  P.~G\'erard, P.~Germain and L.~Thomann,
  {\em On the cubic lowest Landau level equation,}  Arch.\ Rational\ Mech.\ Anal. {\bf  231} (2019) 1073
  \arXiv{1709.04276} [math.AP].

\bibitem{fetter}A.~L.~Fetter, \emph{Rotating trapped Bose-Einstein condensates}, Rev. Mod. Phys. {\bf 81} (2009) 647  \arXiv{0801.2952} [cond-mat.stat-mech].

\bibitem{AO} A.~Biasi, P.~Bizo\'n and O.~Evnin, {\em Solvable cubic resonant systems}, \arXiv{1805.03634} [nlin.SI].

\bibitem{meloturb}S.~Dartois, O.~Evnin, L.~Lionni, V.~Rivasseau and G.~Valette,
  \emph{Melonic turbulence,}
  \arXiv{1810.01848} [math-ph].

\bibitem{abdulmagd}A.~A.~Abul-Magd and A.~Y.~Abul-Magd, \emph{Unfolding of the spectrum for chaotic and mixed systems}, Physica\ A {\bf 396} (2014) 185 \arXiv{1311.2419} [cond-mat.stat-mech].

\bibitem{luukko}P.~Luukko, \emph{Spectral analysis and quantum chaos in two-dimensional nanostructures}, Doctoral dissertation at the University of Jyv\"askyl\"a (2015)\\ \href{https://jyx.jyu.fi/handle/123456789/48270}{https://jyx.jyu.fi/handle/123456789/48270}.

\bibitem{BR} P.~Bizo\'n and A.~Rostworowski,
 {\em On weakly turbulent instability of anti-de Sitter space,}
 Phys.\ Rev.\ Lett.\ {\bf 107} (2011) 031102
 \arXiv{1104.3702} [gr-qc].

\bibitem{rev2} B.~Craps and O.~Evnin,
 {\em AdS (in)stability: an analytic approach,}
Fortsch.\ Phys.\ {\bf 64} (2016) 336
 \arXiv{1510.07836} [gr-qc].

\bibitem{lagpos}T.~Koornwinder, \emph{Positivity proofs for linearization and connection coefficients of orthogonal polynomials satisfying an addition formula,}  J.\ London\ Math.\ Soc.\ {\bf 18} (1978) 101.

\bibitem{lagcomb1}R.~Askey, M.~E.~H.~Ismail and T.~Koornwinder, \emph{Weighted permutation problems and Laguerre polynomials,} J.\ Comb.\ Theor.\ A {\bf 25} (1978) 277.

\bibitem{lagcomb2}D.~Foata and D.~Zeilberger, \emph{Laguerre polynomials, weighted derangements, and positivity,} SIAM\ J.\ Disc.\ Math. {\bf 1} (1988) 425.

\bibitem{GG}P.~G\'erard and S.~Grellier,
{\em The cubic Szeg\H o equation,} Ann. Scient. \'Ec. Norm. Sup. {\bf 43} (2010) 761
\arXiv{0906.4540} [math.CV].

\bibitem{fennell}J.~Fennell, {\em Resonant Hamiltonian systems associated to the one-dimensional nonlinear Schr\"odinger equation with harmonic trapping}, \arXiv{1804.08190} [math.AP].

\bibitem{quintic}A.~Biasi, P.~Bizo\'n and O.~Evnin, {\em Complex plane representations and stationary states in cubic and quintic resonant systems,} \arXiv{1904.09575} [math-ph].

\bibitem{QGP}
  J.~Haegeman, D.~Draxler, V.~Stojevic, J.~I.~Cirac, T.~J.~Osborne and F.~Verstraete,
  \emph{Quantum Gross-Pitaevskii equation,}
  SciPost Phys.\  {\bf 3} (2017) 006
  \arXiv{1501.06575} [quant-ph].

\bibitem{strnorm1}P.~A.~Tomas, \emph{A restriction theorem for the Fourier transform,} Bull.\ Am.\ Math.\ Soc.\ {\bf 81} (1975) 477.

\bibitem{strnorm2}R.~S.~Strichartz, \emph{Restrictions of Fourier transforms to quadratic surfaces and decay of solutions of wave equations,} Duke\ Math.\ J.\ {\bf 44} (1977) 705.

\bibitem{strnorm3}D.~Foschi, \emph{Maximizers for the Strichartz inequality,} J.\ Eur.\ Math.\ Soc.\ {\bf 9} (2007) 739 \arXiv{math/0404011}.

\bibitem{strnorm4}D.~Hundertmark and V.~Zharnitsky, {\em On sharp Strichartz inequalities in low dimensions,} Int.\ Math.\ Res.\ Not.\ {\bf 2006} (2006) 34080,

\end{thebibliography}
\end{document}